\documentclass[11pt, oneside, hidelinks]{article}

\usepackage[letterpaper,margin=1in]{geometry} 

\usepackage{amsmath,amsthm,amssymb, graphicx, multicol, multirow, array,hyperref, bbm, comment, subcaption,  float, booktabs, xpatch, color}
\usepackage{setspace} 

\usepackage{algorithm}
\usepackage{algorithmicx}
\usepackage{algpseudocode}

\def\*#1{\mathbf{#1}}
\def\+#1{\mathbb{#1}}
\def\wt#1{\widetilde{#1}}

\newcommand{\tr}{\mathrm{trace}}

\newcommand{\diag}{\mathrm{diag}}

\newcommand{\RNum}[1]{\uppercase\expandafter{\romannumeral #1\relax}}

\usepackage[T1]{fontenc}
\usepackage[utf8]{inputenc}
\usepackage{lmodern}
\usepackage[english]{babel}
\usepackage[autostyle]{csquotes}
\usepackage{amsmath}
\usepackage{natbib}
\setcitestyle{open={(},close={)}}

\usepackage{hyperref} 
\usepackage{xr}
\makeatletter
\newcommand*{\addFileDependency}[1]{
	\typeout{(#1)}
	\@addtofilelist{#1}
	\IfFileExists{#1}{}{\typeout{No file #1.}}
}
\makeatother

\usepackage{enumitem}
\setlist{itemsep=.01em}
\setlist{topsep=.5em}

\newtheorem{innercustomgeneric}{\customgenericname}
\providecommand{\customgenericname}{}

\newtheorem{theorem}{Theorem}
\newtheorem{lemma}{Lemma}
\newtheorem{proposition}{Proposition}
\newtheorem{remark}{Remark}

\newtheorem{definition}{Definition}
\newtheorem{assumption}{{Assumption}}

\def\beq{\begin{equation}}
\def\eeq{\end{equation}}
\def\beqr{\begin{eqnarray}}
\def\eeqr{\end{eqnarray}}
\def\beqrs{\begin{eqnarray*}}
\def\eeqrs{\end{eqnarray*}}
\def\bet{\begin{theorem}}
\def\eet{\end{theorem}}
\def\bel{\begin{lemma}}
\def\eel{\end{lemma}}
\def\bep{\begin{proposition}}
\def\eep{\end{proposition}}
\def\bg{\begin{figure}[tbph]\begin{center}}
\def\eg{\end{center}\end{figure}}

\def\bc{\begin{center}}
\def\ec{\end{center}}

\def\wt{\widetilde}
\def\wh{\widehat}

\def\argmin{\mbox{argmin}}

\def\diag{\mbox{diag}}

\newcommand{\Var}{\textnormal{Var}}
\newcommand{\Cov}{\textnormal{Cov}}

\newcommand{\bB}{{\mathbf B}}

\newcommand{\bG}{{\mathbf G}}
\newcommand{\bH}{{\mathbf H}}
\newcommand{\bI}{{\mathbf I}}
\newcommand{\bK}{{\mathbf K}}

\newcommand{\bM}{{\mathbf M}}

\newcommand{\bR}{{\mathbf R}}
\newcommand{\bS}{{\mathbf S}}

\newcommand{\bU}{{\mathbf U}}
\newcommand{\bV}{{\mathbf V}}
\newcommand{\bW}{{\mathbf W}}
\newcommand{\bX}{{\mathbf X}}
\newcommand{\bY}{{\mathbf Y}}

\newcommand{\ba}{{\mathbf a}}
\newcommand{\bb}{{\mathbf b}}

\newcommand{\be}{{\mathbf e}}
\newcommand{\bff}{{\mathbf f}}
 
\newcommand{\bh}{{\mathbf h}}

\newcommand{\bp}{{\mathbf p}}

\newcommand{\br}{{\mathbf r}}

\newcommand{\bu}{{\mathbf u}}
\newcommand{\bv}{{\mathbf v}}
\newcommand{\bw}{{\mathbf w}}

\newcommand{\by}{{\mathbf y}}

\newcommand{\bbeta}  {\boldsymbol{\beta}}
\newcommand{\bfeta}  {\boldsymbol{\eta}}

\newcommand{\bdelta} {\boldsymbol{\delta}}

\newcommand{\bOmega}{\boldsymbol{\Omega}}

\newcommand{\bSigma}{\boldsymbol{\Sigma}}

\newcommand{\bve}{\mbox{\boldmath$\varepsilon$}}
\newcommand{\brho}{\mbox{\boldmath$\rho$}}

\newcommand{\btheta} {\boldsymbol{\theta}}
\newcommand{\bxi} {\boldsymbol{\xi}}

\newcommand{\bmu} {\boldsymbol{\mu}}
\newcommand{\bzeta} {\boldsymbol{\zeta}}
\newcommand{\bGamma} {\boldsymbol{\Gamma}}
\newcommand{\bLambda} {\boldsymbol{\Lambda}}

\newcommand{\bD}{{\mathbf D}}

\newcommand{\bnu}{\boldsymbol{\nu}}

\newcommand{\ve}{{\varepsilon}}
\renewcommand{\epsilon}{{\ve}}
\renewcommand{\hat}{\widehat}

\def\wt{\widetilde}
\renewcommand{\tilde}{\wt}

\newcolumntype{L}[1]{>{\raggedright\arraybackslash}p{#1}}

\usepackage[margin=10pt,labelfont=bf]{caption}

\begin{document}
	


\title{\Large High-Dimensional Spatial Arbitrage Pricing Theory with Heterogeneous Interactions}
\date{}
	\author{
Zhaoxing Gao$^1$,  Sihan Tu$^2$, and Ruey Tsay$^3$\thanks{Corresponding author: \href{mailto:ruey.tsay@chicagobooth.edu}{ruey.tsay@chicagobooth.edu} (R.S. Tsay).  Booth School of Business, University of Chicago, 5807 South Woodlawn Avenue, Chicago, IL, 60637, USA.} \\ 
{\normalsize $^1$School of Mathematical Sciences, University of Electronic Science and Technology of China}\\ {\normalsize $^2$School of Management, Zhejiang University}\\
{\normalsize $^3$Booth School of Business, University of Chicago}
}
	
	\begin{onehalfspacing}
		\begin{titlepage}
		\maketitle
			\vspace{-1cm}
\begin{abstract}
This paper investigates
estimation and inference of a Spatial Arbitrage Pricing Theory (SAPT) model that integrates spatial interactions with multi-factor analysis, accommodating both observable and latent factors.
Building on the classical mean-variance analysis, we 
introduce a class of Spatial Capital Asset Pricing Models (SCAPM) that account for spatial effects in high-dimensional assets, where we define {\it spatial rho} as a counterpart to market beta in CAPM. We then extend SCAPM to a general SAPT framework under a {\it complete} market setting by incorporating multiple factors.
For SAPT with observable factors, we propose a generalized shrinkage Yule-Walker (SYW) estimation method that integrates ridge regression to estimate spatial and factor coefficients. When factors are latent, we first apply an autocovariance-based eigenanalysis to extract factors, 
then employ the SYW 
method using the estimated factors. We establish asymptotic properties for these estimators under high-dimensional settings where both the dimension and sample size diverge. 
Finally, we use simulated and real data examples to demonstrate 
the efficacy and usefulness of the proposed model and method. 


				\vspace{1cm}
				
				
			\end{abstract}

		\end{titlepage}

\section{Introduction}
With the rapid advancement in information technology, large-scale datasets have become ubiquitous across 
all scientific areas with important applications. These datasets also introduce new analytical challenges in financial econometrics and statistics, particularly in high-dimensional settings.
As a fundamental tool for dimension reduction and feature extraction, factor models 
provide a crucial link between economic theory and data analysis.
Since the seminal work of \cite{markowits1952portfolio} on portfolio theory, factor-based pricing models have played a central role in asset pricing, investment analysis and risk assessment. The Capital Asset Pricing Model (CAPM), developed by Sharpe, Lintner, and Mossin in the 1960s, introduced the concept of {\it market beta} to quantify systematic risk-return relationships.  \cite{Ross1976} proposed the Arbitrage Pricing Theory (APT), which extended the single-factor CAPM by incorporating multiple systematic risk factors under no-arbitrage principles, allowing for a more flexible representation of expected returns.
Modern factor-based pricing research has evolved into two dominant approaches to address 
the growing market complexity. The first  approach, developed by  \cite{fama1993common, fama2015five}, relies on the theory-driven observable factors, such as market returns and firm characteristics. Building on this framework, numerous factor models for asset returns have been proposed; for instance, \citet{feng2020taming} propose the Double-Selection LASSO to evaluate the marginal contribution of individual factors relative to an existing high-dimensional factor set.
While these models offer strong economic interpretability, their fixed factor structures limit their ability to capture modern dynamic market interactions.
Recent studies by \cite{forni2000generalized}, \cite{bai2002determining}, \cite{bai2003inferential}, \cite{forni2005generalized}, \cite{lam2012factor}, \cite{fan2013large}, \cite{gao2022modeling,gao2023divide}, among others, have focused on latent factor models as an alternative approach. These models provide a methodology for inferring unobserved common factors from covariance structures. \citet{lettaupelger2019} and \citet{giglio2025test} further demonstrate the effectiveness of their tailored latent factor models in asset pricing, offering deeper insights into the underlying structure of financial markets. \cite{liuguerardchentsay} 
show that one can improve the 
estimation of portfolio risk by augmenting the Fama and French factors with latent factors extracted from 
a matrix-variate dataset of asset returns.

Despite the effectiveness of factor models in explaining cross-sectional and dynamic dependence, many economic and financial applications often manifest intricate spatial interconnections. 
Consider, for example, the spatial distribution of economic indicators across regions, where the performance of one region may influence its neighbors; see \cite{anselin1988spatial} and \cite{cressie2015statistics}. Since the seminal work of \cite{cliff1973} on spatial autocorrelations, spatial models are often used to model cross-sectional dependence of different economic units or individuals at different locations. 
More recently, the spatial models have been extended to spatial dynamic panel data (SDPD) models by adding a time-lagged direction to account for serial correlations across different economic units or individuals; see, for example, \cite{lee2010some}. Empirically, the spatial interactions among the panel may exist in many large-dimensional economic and financial systems, together with other comovements or common factors. For example, \cite{pirinsky2006does} found the spatial effect in the U.S. equity market by studying the comovements of common stock returns of U.S. corporations in the same geographic area; \cite{kou2018asset} proposed an asset pricing model with spatial interactions and discovered significant spatial interactions in the futures contracts on S\&P/Case-Shiller Home Price Indices. Therefore, augmenting factor models with spatial interactions not only extends these models with additional common factors but also enriches spatio-temporal models by integrating common factor structures. 

In this paper, we focus on spatial panel models with common factors in the context of arbitrage pricing under high-dimensional settings. Building on the classical mean-variance analysis, we first introduce a class of Spatial Capital Asset Pricing Models (SCAPM) that account for spatial effects in high-dimensional assets under a ``{\it complete market}'' or ``{\it minimum complete market}'' assumption, where we introduce a {\it spatial rho} as a counterpart to market beta in CAPM. 
Within the spatial CAPM framework, we extend the model to a Spatial Arbitrage Pricing Theory (SAPT) by incorporating a multifactor structure. This formulation captures both systematic risk factors and spatial spillover effects, offering a unified approach to modeling interdependencies in asset returns.

While prior studies, such as \cite{pesaran2011large}, \cite{kou2018asset}, \cite{bai2021dynamic}, \cite{yang2021common}, and \cite{hu2023arbitrage}, have examined similar spatial interactions in factor models, the SAPT studied in this paper differs from the existing models for several reasons.
First, unlike \cite{pesaran2011large}, which focuses on spatial autocorrelation in unobserved errors, our model explicitly captures spatial correlations among panel units. Second, the proposed SAPT model functions as a pure spatial arbitrage pricing factor model without lagged or exogenous variables, distinguishing it from the models in \cite{bai2021dynamic} and \cite{yang2021common}, which incorporate exogenous features and assume a homogeneous spatial coefficient. This structure presents challenges in identifying suitable instrumental variables for method-of-moments estimation.
Third, we consider both observable and latent factor structures. When factors are observable, our model aligns with the spatial asset pricing models of \cite{kou2018asset} and \cite{hu2023arbitrage} for financial returns. However, when factors are unobservable, which is not considered in \cite{kou2018asset} or \cite{hu2023arbitrage},  our model extends the statistical and econometric factor models by incorporating spatial interaction terms, capturing additional panel information beyond common latent factors.
Fourth, our model accommodates panel dimensions that can grow to infinity, differing from the quasi-maximum likelihood estimation (QMLE) framework in \cite{aquaro2021estimation} and \cite{hu2023arbitrage}, where the dimension is fixed. This flexibility enables broader applications in high-dimensional settings.

These distinctive features of the proposed SAPT model introduce additional estimation challenges, making conventional spatial econometric methods inadequate.
For models with observable factors, the widely used QMLE approach, as discussed in \cite{lee2004asymptotic}, \cite{yu2008quasi}, and \cite{bai2021dynamic}, often encounters computational difficulties due to the large matrix determinants involved in the likelihood function. These challenges become even more pronounced in high-dimensional settings, especially when estimating numerous unit-specific spatial coefficients.
In cases with heteroskedastic disturbances, \cite{lin2010gmm} demonstrated that the QML estimator for the spatial autoregressive (SAR) model is inconsistent if heteroskedasticity is ignored. To address this problem, they proposed a GMM estimator, which is computationally more efficient than QMLE. However, the SAPT model considered here lacks lagged or exogenous variables, making it difficult to identify suitable instrumental variables for constructing sufficiently many estimating equations.


In view of this, we propose a ridge-regularized Yule-Walker estimator that integrates shrinkage techniques with method-of-moments. By incorporating lagged common factors as instrumental variables, we reformulate parameter estimation as a system of $L_2$-penalty Yule-Walker equations for each panel component, thereby addressing the issue of insufficient number of estimating equations in settings without exogenous variables or structural constraints on spatial effects. 
{In contrast to the regularized method-of-moments approaches proposed by \citet{liao2013adaptive} and \citet{carrasco2015regularized}, which primarily focus on selecting instruments or moment conditions,  our method applies ridge regularization directly to the Yule-Walker equations to mitigate potential singularity and improve estimation robustness.}  We establish the asymptotic properties of our estimator in the setting where both the dimension $N$ and the sample size $T$ approach infinity. Despite the bias inherent in ridge estimators, we demonstrate the feasibility of conducting joint parameter inference. This contrasts with QML estimators in \cite{aquaro2021estimation} and \cite{hu2023arbitrage} which often require finite $N$ and inevitably accumulate asymptotic bias as $N$ diverges with $T$; See Remarks 6 and 7 in \cite{aquaro2021estimation} for a discussion.
While alternative methods, such as those proposed by \cite{bai2021dynamic}, ensure parameter consistency under structural constraints and complex bias correction, it remains unclear whether their approach is feasible for the SAPT considered in this paper with heterogeneous spatial interactions. 

In the presence of latent factors, 
our model can be reformulated as an approximate factor model. We propose a two-step procedure to extract latent factors and to estimate unknown parameters. 
Given the white noise assumption on the error terms in the SAPT model (see \cite{kou2018asset}, \cite{aquaro2021estimation}, and \cite{hu2023arbitrage}), we first apply the auto-covariance-based eigenanalysis approach from \cite{lam2012factor} and \cite{gao2022modeling} to estimate dynamically dependent factors. This ensures that the factors and their lagged counterparts remain uncorrelated with the noise terms, enabling their use as instrumental variables.
Once the factors are extracted via eigenanalysis, we implement the Yule-Walker estimation method, replacing the unknown factors with their estimated counterparts. Furthermore, we establish the asymptotic properties of the estimated factors, scalar coefficients, and loading vectors as both the dimension
$N$ and sample size 
$T$ approach infinity. Notably, we also derive the limiting distributions of the estimated factors under a proper rotation matrix, a result not presented in \cite{lam2012factor}, offering independent interest for readers.

{We conduct extensive simulations to evaluate the accuracy of our estimation method, particularly in estimating the {\it spatial rho} and the loading matrix, while examining the convergence and asymptotic properties of the jointly estimated parameters.
Moreover, we compare our method’s predictive performance with QML estimators. The results show that our approach outperforms these alternatives in out-of-sample forecasting. Empirically, we apply our method to two real datasets on U.S. stock returns and housing prices, respectively. In both cases, it achieves superior out-of-sample forecasting performance compared to QMLE and the classical Fama-French factor model, reinforcing its practical advantages in high-dimensional economic and financial analysis.}

This paper makes several significant contributions. First, rather than relying on a mathematical formulation, we derive the SCAPM from a classical mean-variance perspective and extend it to a SAPT framework by integrating a multifactor structure. This approach offers a new perspective for economists and practitioners in understanding spatial asset pricing theory.
Second, from a modeling standpoint, the proposed framework is flexible, accommodating both observable and latent factors. This extension provides an opportunity to explore the dynamics of large-dimensional economic and financial panel systems.
Third, from a methodological perspective, since QMLE methods require extensive computation and may be impractical in high-dimensional settings with general covariance structures, we propose a shrinkage estimation approach with joint inferential theory for the proposed models. While individual estimators may not be consistent, joint estimation allows for consistent inference. Our procedure is computationally efficient and avoids the need for restrictive distributional or covariance assumptions when using the Yule-Walker estimation method. More importantly, the proposed shrinkage estimation method  outperforms the QMLE method in out-of-sample evaluations, highlighting their empirical advantages in high-dimensional applications.

The remainder of the paper is structured as follows. Section~\ref{sec20} outlines the formulations of the SCAPM and SAPT models under study. Section \ref{sec2} provides the modeling framework and its estimation procedure. Section \ref{sec3} establishes the asymptotic properties of the derived estimators. Section \ref{sec4} evaluates the finite-sample performance of the proposed approach through simulations and Section \ref{sec5}  illustrates the proposed model and method with two empirical applications. Section 5 concludes.   All 
proofs and derivations for the asymptotic results are relegated to an online Appendix.


{\bf Notation:}  We use the following notation. For a $p\times 1$ vector
$\bu=(u_1,..., u_p)'$, $\|\bu\|_1=\sum_{i=1}^p|u_i|$ is the $\ell_1$-norm and $\|\bu\|_\infty=\max_{1\leq i\leq p}|u_i|$ is the $\ell_\infty$-norm. $\bI_p$ denotes the $p\times p$ identity matrix. For a matrix $\bH$, its Frobenius norm is $\|\bH\|=[\tr(\bH'\bH)]^{1/2}$ and its operator norm is $\|\bH
\|_2=\sqrt{\lambda_{\max} (\bH' \bH ) }$, where
$\lambda_{\max} (\cdot) $ denotes the largest eigenvalue of a matrix, and $\|\bH\|_{\min}$ is the square root of the minimum non-zero eigenvalue of $\bH\bH'$. $|\bH|$ denotes the absolute value of $\bH$ elementwisely. The superscript ${'}$ denotes the 
transpose of a vector or matrix. We also use the notation $a\asymp b$ to denote $a=O(b)$ and $b=O(a)$ or $a$ and $b$ have the same order of stochastic bound when they are random variables.

\section{Spatial CAMP and Spatial APT}\label{sec20}
In this section, we develop a Spatial Capital Asset Pricing Model (SCAPM) using mean-variance analysis within a {\it complete market} framework. Additionally, we construct a spatial arbitrage pricing theory model by incorporating a multifactor structure.

\subsection{Complete Market Assumption}
We consider a one-period economy with $N$ risky assets in the market whose random returns are denoted as $\br=(r_{1},...,r_{N})'$ over the period.
The expected return is $\bmu=(\mu_1,...,\mu_N)'$ and the risk-free asset return is $r_f$. Let $r_{M}$ be the return of the market portfolio or the tangency portfolio in the mean-variance framework of \cite{markowits1952portfolio} with expected return $\mu_M$.

Suppose the $N$ assets are diverse, with $N$ sufficiently large, and encompass a wide spectrum of asset categories.  It is reasonable to assume that each individual asset exhibits some degree of association with the others. Within this framework, we introduce the concept of a {\it complete market}, which implies that the extensive set of assets enables the formation of suitable linear combinations to replicate the returns of any specific asset in the market. The definition of a {\it complete market}, or equivalently, a {\it minimum complete market}, is provided in Definition~\ref{def1} and equivalently in Definition~\ref{def2} below.


\begin{definition}[{\it Complete Market}]\label{def1}
   Suppose there are 
$N$ risky assets in the market, where 
$N$ is sufficiently large. The market is said to be complete if the return of any asset 
$r_j$  can be expressed as a linear combination of the remaining 
$N-1$ assets, i.e., those indexed by $\{1,...,N\}\setminus\{j\}$, for $j=1,...,N$, However, 
$r_j$ cannot be replicated using only 
$N-2$ assets  from $\{1,...,N\}\setminus\{j\}$. 
\end{definition}

\begin{definition}[{\it Minimum Complete Market}]\label{def2}
A market is called a minimum complete market if it contains at least 
$N-1$ assets from a complete market, as the return of the remaining asset can be fully replicated by a linear combination of the other 
$N-1$ assets in  this 
minimum complete market.
\end{definition}

This conceptualization of a complete market aligns with the idea that the abundance and diversity of assets enable the construction of portfolios capable of replicating the performance of any individual asset. It suggests that the richness of the market, in terms of asset variety, allows for the creation of synthetic versions of assets by leveraging a diverse set of available instruments.

In a high-dimensional setting, market completeness arises from the vast number and diversity of assets, facilitating the construction of well-diversified portfolios that can closely approximate the returns of specific assets. The concept of a complete market is closely tied to the absence of arbitrage opportunities. In such a market, no-arbitrage conditions ensure that riskless profits cannot be generated through linear combinations of available assets. If arbitrage opportunities existed, they would indicate an incomplete market, as investors could exploit them to create new assets beyond those initially available.

However, achieving a truly complete market in practice is challenging. Real-world markets often face limitations in asset variety, and factors such as transaction costs, market frictions, and short-selling constraints can hinder perfect asset replication. Nonetheless, the notion of a complete market provides a framework for understanding the relationships among assets and their pricing dynamics in a diversified financial environment.

\subsection{From CAPM to Spatial CAPM}\label{scapm:in}
\begin{figure}[htp]
\begin{center}
{\includegraphics[width=0.6\textwidth]{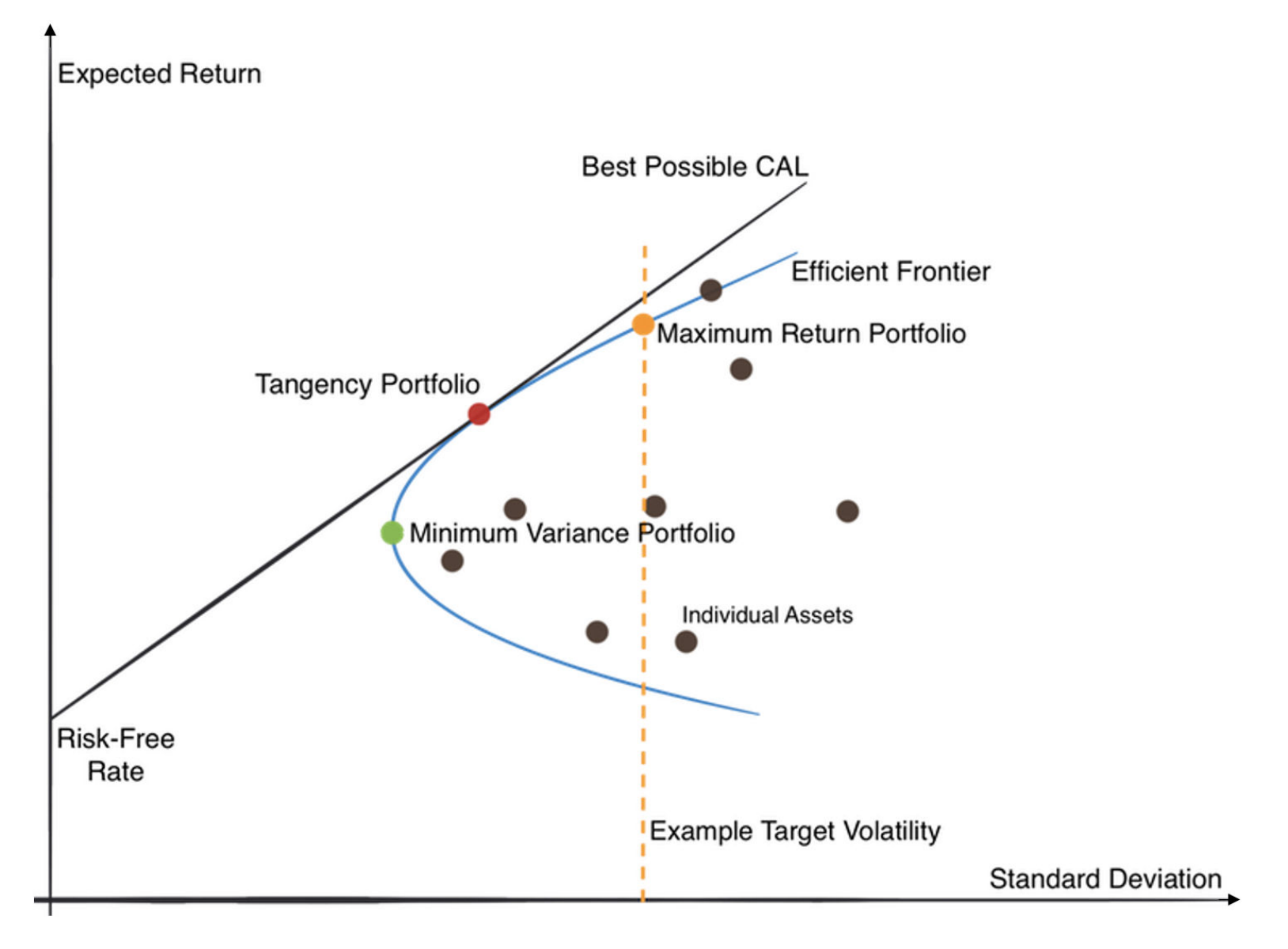}}
\caption{Mean-variance efficient frontier with a risk-free asset. The horizontal axis denotes the standard deviation of the portfolio and the vertical axis denotes the expected return of the corresponding portfolio. Available at \url{https://quantpedia.com/markowitz-model/}.}\label{fig00}
\end{center}
\end{figure}
Based on the mean-variance analysis  (e.g., \cite{cochrane2009asset}), there exists a weight vector $\btheta=(\theta_1,...,\theta_N)'$ such that the market (or tangency) portfolio can be expressed as 
 $r_{M}=\btheta'\br$, as illustrated in Figure~\ref{fig00}. For the $j$-th asset with return $r_{j}$ and expected return $\mu_j$, the capital asset pricing model (CAPM) of \cite{sharpe1964capital} states that
\[\mu_j-r_f=\frac{\Cov(r_{j},r_{M})}{\Var(r_{M})}(\mu_M-r_f),\]
where the quantity $\beta_j=\frac{\Cov(r_{j,t},r_{M,t})}{\Var(r_{M,t})}$  is  referred to as the market beta of the $j$th asset in the finance literature. In practice,  the S\&P 500 index return often 
serves as a proxy for the market portfolio, and the market beta can be estimated by running an OLS regression over $T$ periods. For further details, see Chapters 5 and 9 of \cite{cochrane2009asset}.



Next, we formulate a spatial capital asset pricing model (SCAPM), building on the mean-variance analysis within a complete market defined in Definition~\ref{def1}.
 For each  $j$, we remove $r_{j}$ from the portfolio return vector $\br$ and consider the mean-variance analysis of the remaining $N-1$ risky assets and the risk-free rate $r_f$. Through 
 the classic mean-variance optimization, we obtain the portfolio weight $\bw_j$, where the $j$-th position of $\bw_j$ is zero and $\bw_j'{\bf 1}_N=1$ such that the portfolio $\bw_j'\br$ is a tangency portfolio, as illustrated in Figure~\ref{fig00} without the $j$-th asset. The optimal portfolios lie along the capital allocation line (CAL) in the mean-variance framework, with a slope 
\[\frac{\mu_{j,M}-r_f}{\sigma_{j,M}},\] where
\[\mu_{j,M}=E(\bw_j'\br),\,\,\text{and}\,\, \sigma_{j,M}=\sqrt{\Var(\bw_j'\br}).\]
Then, for the asset $j$ with expected return $\mu_j$, we have the following theorem.
\begin{theorem}\label{thm0}
   Suppose the 
$N$ risky assets are in a complete market, as described in Definition~\ref{def1}. For the 
$j$-th risky asset with expected return $\mu_j$, we have the following relationship:
    \[\mu_j-r_f=\frac{\Cov(r_{j},\bw_{j}'\br)}{\Var(\bw_j'\br)}(\mu_{j,M}-r_f),\]
    where $\mu_{j,M}$ is the expected return of the tangency portfolio $r_{j,M}=\bw_j'\br$ with the $j$-th asset  excluded from the portfolio. We define $\rho_j=\frac{\Cov(r_{j},\bw_{j}'\br)}{\Var(\bw_j'\br)}$ and refer to it  
    as the ``spatial rho'' for the $j$-th asset.
\end{theorem}

The proof of Theorem~\ref{thm0} can be found in the Appendix. From Theorem~\ref{thm0}, we observe that the spatial rho is asset-specific, similar to the market beta in the CAPM. However, the key difference is that the  spatial tangent portfolio is also asset-specific, which contrasts with the classical CAPM, where the market portfolio is fixed and unique for all assets.

\subsection{Spatial Arbitrage Pricing Theory}\label{sapt:in}
In this section,  we derive a spatial  arbitrage pricing theory model following the framework of \cite{Ross1976}. To better illustrate the application of the proposed model in asset pricing, we use the notation $\br_t=(r_{1,t},...,r_{N,t})'$ to denote a vector of returns to $N$ risky assets at time $t$. Letting $\bmu_0=(\mu_{0,1},...,\mu_{0,N})'$ be the expected returns of $\br_t$, we consider the following asset pricing model with spatial interactions and multi-factors:
\begin{align}\label{apt:s}
    \br_t &=\bD(\boldsymbol{\rho})\bW\br_t+\bnu_0+\bB\bff_t+\bve_t,
\end{align}
where $\bff_t=(f_{1,t}$,...,$f_{K,t})'$ consists of $K$ observable factors for which the expected return 
of $f_{i,t}$ is $\mu_i$, for $i=1,...,K$. The columns of $\bB=(\bdelta_1,...,\bdelta_K)$  are the associated $K$ loading vectors of the $K$ factors, and $\bnu_0=(\bI_N-\bD(\boldsymbol{\rho})\bW)\bmu_0$, where $\boldsymbol{\rho}$ is a vector of spatial rhos.
$\bW$ is a known spatial weight matrix with zero main diagonal elements, and $\bD(\boldsymbol{\rho})= diag(\rho_1,...,\rho_N)$, where $\rho_j$ can be estimated by the method in Section~\ref{sec2} below. We may assume each row of $\bW$, denoted as $\bw_j$, can be calculated either based on some economic distance or through 
the mean-variance analysis.
We introduce some notations before the derivation of the spatial arbitrage pricing theory. We use ${\bf 1}_N$ to denote the $N$-dimensional vector of 1, e.g., ${\bf 1}_N=(1,...,1)'\in R^N$. Let $\btheta=(\theta_1,...,\theta_N)'$ represent the weight vector that will be used to construct an arbitrage portfolio. Our derivation proceeds in the following three steps.\\

{\noindent \it Step 1.}
Suppose the random vector of returns $\br_t$ satisfies Model (\ref{apt:s}). We use a weight vector $\btheta$ to construct an arbitrage portfolio of $N$ assets, where we assume  $\btheta'{\bf 1}_N=0$, implying that there is no wealth invested in the portfolio. We also 
require $\btheta$ to be a well-diversified portfolio weight with each component $\theta_i$ being of order $1/N$ in  magnitude as in \cite{Ross1976}.\\

{\noindent\it Step 2.} The random return of the portfolio can be written as
\[\btheta'\br_t=\btheta'\bmu_0+\btheta'(\bI_N-\bD(\boldsymbol{\rho})\bW)^{-1}\bB\bff_t+\btheta'(\bI_N-\bD(\boldsymbol{\rho})\bW)^{-1}\bve_t,\]
 where $\bmu_0=(\bI_N-\bD(\boldsymbol{\rho})\bW)^{-1}\bnu_0$. We further assume that $\ve_{i,t}$'s are independent with each other, for $i$ and $t$, which is a commonly used assumption in the spatial econometrics literature, and each element of $\bS(\boldsymbol{\rho})^{-1}=(\bI_N-\bD(\boldsymbol{\rho})\bW)^{-1}$ are of order $1/N$ in absolute magnitude. Together with Assumption~\ref{asm2} in Section~\ref{sec3} below, by the law of large numbers, we can show that
 \[\btheta'(\bI_N-\bD(\boldsymbol{\rho})\bW)^{-1}\bve_t=o_p(1),\]
 and, hence, 
 \[\btheta'\br_t\approx\btheta'\bmu_0+\btheta'(\bI_N-\bD(\boldsymbol{\rho})\bW)^{-1}\bB\bff_t.\]

 {\noindent \it Step 3.} If we require that the arbitrage portfolio with weight $\btheta$ be chosen with no systematic risk, then
 \begin{equation}\label{theta:b}
  \btheta'(\bI_N-\bD(\boldsymbol{\rho})\bW)^{-1}\bdelta_i=0,\quad i=1,...,K.    
 \end{equation}
 This condition ensures that the return of the arbitrage portfolio becomes $\btheta'\bmu_0$. Using the constraint of no wealth that $\btheta'{\bf 1}_N=0$, the return must be zero to prevent arbitrarily large disequilibrium positions. Therefore, we have
 \begin{equation}\label{theta:mu}
  \btheta'\bmu_0=0.
 \end{equation}
 From the relationships in (\ref{theta:b}), (\ref{theta:mu}), and $\btheta'{\bf 1}_N=0$, we conclude that $\bmu_0$, ${\bf 1}_N$, and $(\bI_N-\bD(\boldsymbol{\rho})\bW)^{-1}\bdelta_i$ are on the same hyperplane, for $i=1,...,K$. Then there exist $\gamma_0$, $\gamma_1$,..., $\gamma_K$ such that
 \begin{equation}\label{mu0}
 \bmu_0=\gamma_{0,i}{\bf 1}_N+\gamma_i(\bI_N-\bD(\boldsymbol{\rho})\bW)^{-1}\bdelta_i,\quad i=1,...,K.    
 \end{equation}
 We will solve the above equations for $\gamma_{0,i}$ and $\gamma_i$ by a plug-in method. 
 Note that when $\bmu_0=r_f {\bf 1}_N$, the return vector of a risk-free asset, the loadings associated with the factors are zero, i.e.,  $\bdelta_i={\bf 0}$, for $i=1,...,K$. Furthermore, if we take $\br_t=f_{i,t}{\bf 1}_N$, then $\bmu_0=\mu_{i}{\bf 1}_N$, and the spatial parameter $\boldsymbol{\rho}={\bf 0}$, since there is no spatial effect for a single asset.  In this case, the exposure to the $i$-th factor is $\bdelta_i={\bf 1}_N$, while the exposures to the other factors are zero. 
 These special cases result in the following equations:
\[\left\{\begin{array}{cc}
     r_f{\bf 1}_N=\gamma_{0,i} {\bf 1}_N,  \\
     \mu_i{\bf 1}_N=\gamma_{0,i} {\bf 1}_N+\gamma_i{\bf 1}_N,\quad i=1,...,K.
\end{array}\right.\]
It follows from the above equations that
\[\gamma_{0,i}=r_f,\,\, \gamma_i=\mu_i-r_f,\quad i=1,...,K,\]
where $\gamma_{0,i}$ turns out to be independent of $i$.
Then, (\ref{mu0}) becomes
\[\bmu_0=r_f{\bf 1}_N+(\bI_N-\bD(\boldsymbol{\rho})\bW)^{-1}\bdelta_1(\mu_1-r_f)+...+(\bI_N-\bD(\boldsymbol{\rho})\bW)^{-1}\bdelta_K(\mu_K-r_f),\]
or equivalently, 
\[(\bI_N-\bD(\boldsymbol{\rho})\bW)(\bmu_0-r_f{\bf 1}_N)=\bdelta_1(\mu_1-r_f)+...+\bdelta_K(\mu_K-r_f),\]
which is a spatial APT model that extends the SCAMP in Section~\ref{scapm:in} with multi-factors, where $\mu_i-r_f$ is the risk premium of the $i$-th factor and $\bmu_0-r_f{\bf 1}_N$ is the vector of $N$ excessive asset returns. For the $j$-th asset, we can derive that
\begin{equation}\label{scapm}
 \mu_{0,j}-r_f=\rho_j\bw_j'(\bmu_0-r_f{\bf 1}_N)+\bdelta_{1,j}(\mu_1-r_f)+...+\bdelta_{K,j}(\mu_K-r_f),\,\,j=1,...,N.
\end{equation}
Therefore, we may construct a new asset-specific factor, called the spatial factor, defined as $\bw_j'(\bmu_0-r_f{\bf 1}_N)$ associated with the $j$-th asset where the $j$-th element of $\bw_j$ is zero according to the definition of the spatial weight. The scalar $\rho_j$ represents the spatial effect on the 
$j$-th asset, which is termed the {\it spatial rho}, in contrast to the {\it market beta} in the classic CAPM of \cite{sharpe1964capital}.


In the next section, we examine a general APT model that incorporates spatial interactions and propose a Yule-Walker estimation and inference method using factor instruments and ridge techniques for the model.

\section{General Model and Methodology} \label{sec2}
		
		\subsection{Setup}\label{model_overview}
		Let $\by_t = (y_{1,t}, ..., y_{N,t})'$ be an $N$-dimensional observable panel of time series at time $t$, where we assume all the data are centered with zero mean. Thus, $\by_t$ replaces $(\br_t - \bmu_0)$ in Model (\ref{apt:s}), and the factors $\{\bff_t\}$, for $ t = 1, ..., T$, are assumed to have zero mean. Based on  the SAPT model in Section~\ref{sapt:in}, we assume that $\by_t$ follows the following general structure:
  \begin{equation}\label{ft:sp}
      \by_t=\bD(\boldsymbol{\rho})\bW\by_t+\bB\bff_t+\bve_t,\,\, t=1,...,T,
  \end{equation}
where $\bff_t$ is a $K$-dimensional factor process that is either observable or unobservable, $\bB$ is the loading matrix associated with the factors, $\bW$ is the $N \times N$ spatial weight matrix that measures the dependence among different economic units or individuals of $\by_t$. $\bD(\boldsymbol{\rho}) = \diag(\rho_1, ..., \rho_N)$, where $\rho_j$ is an unknown coefficient parameter for the $j$-th individual. $\bve_t$ is a white noise term that is uncorrelated with $\bff_t$, but we allow for dependence between $\bff_{t+j}$ and $\bve_t$, for $j \geq 1$, since the factors $\bff_t$'s are usually serially dependent, which may be correlated with some lagged noise terms.

It is a common practice in spatial econometrics to assume that $\bW$ is known, and the main diagonal elements of $\bW$ are zero. The weights may be based on physical distance, social networks, or ``economic” distance, as seen in \cite{case1993budget}. For example, we may take $w_{ij} := (s_i d_{ij})^{-1}$, for $i \neq j$,  and $w_{ii} = 0$, where $d_{ij}$ is the physical distance between location $i$ and location $j$, and $s_i := \sum_j d_{ij}^{-1}$. Alternatively, we may take $d_{ij}^{-1}$ as the sample correlation between the $i$-th and $j$-th economic units when there is no clear physical distance between them.
When $\rho_1 = ... = \rho_N$, the spatial interaction term in Model (\ref{ft:sp}) reduces to the classical setting in the spatial econometrics literature, such as \cite{lee2004asymptotic}, among others.

For a given spatial weight matrix $\bW = (\bw_1, \dots, \bw_N)'$, where $\bw_i$ is the $i$-th row vector of $\bW$, our goal is to estimate the unknown coefficients in $\boldsymbol{\rho}$ and $\bB$ when the factors $\bff_t$'s are observable. When the factors $\bff_t$'s are latent, we also need to recover the latent factors.


		\subsection{Shrinkage Yule-Walker Estimation with Observed Factors}\label{sec22}
In this section, we study the scenario when the factors are observed and propose a generalized shrinkage Yule-Walker method to estimate the unknown coefficients, which is essentially a combination of ridge regression and the method-of-moments. To this end, we begin with some useful notation. Define $\bSigma_{yf}(k) = \Cov(\by_t, \bff_{t-k})$ as the covariance matrix between $\by_t$ and the past lagged factor variables $\bff_{t-k}$, and $\bSigma_{f}(k) = \Cov(\bff_t, \bff_{t-k})$ as the lag-$k$ auto-covariance matrix of $\bff_t$, for $k \geq 0$. 
Then, Model (\ref{ft:sp}) implies that
\begin{equation}\label{YW:E}
\bSigma_{yf}(k)=\bD(\boldsymbol{\rho})\bW\bSigma_{yf}(k)+\bB\bSigma_f(k),\quad k\geq 0.    
\end{equation}
Let $\be_i$ be the $i$th unit vector with the $i$th element equal to $1$ and other elements being zero. For each $k\geq 0$, it follows from (\ref{YW:E}) that 
\begin{equation}\label{e:sig}
  \be_i'\bSigma_{yf}(k)=\be_i'\bD(\boldsymbol{\rho})\bW\bSigma_{yf}(k)+\be_i'\bB\bSigma_f(k),\quad i=1,...,N.
\end{equation}
Note that $\be_i'\bD(\boldsymbol{\rho})\bW=\rho_i\bw_i'$ and $\be_i'\bB=\bb_i'$, where $\bw_i$ and $\bb_i$ are the $i$th row vectors of $\bW$ and $\bB$, respectively. Then, (\ref{e:sig}) becomes
\begin{equation}\label{pop:YW}
\bSigma_{yf}'(k)\be_i=\bSigma_{yf}'(k)\bw_i\rho_i+\bSigma_f'(k)\bb_i,\quad i=1,...,N.
\end{equation}
In practice, given the sample data $\{(\by_t,\bff_t):t=1,...,T\}$, by a similar argument to the Yule-Walker estimation method with a given lag $k\geq 0$, we may solve the following minimization problem:
\begin{equation}\label{yw:est}
(\wh\rho_i,\wh\bb_i')'=\arg\min_{\rho\in R,\bb\in R^{r}}\{\|\wh\bSigma_{yf}'(k)\be_i-\wh\bSigma_{yf}'(k)\bw_i\rho-\wh\bSigma_f'(k)\bb\|_2^2\},\,\,i=1,...,N,
\end{equation}
where
\[\wh\bSigma_{yf}(k)=\frac{1}{T}\sum_{t=k+1}^T\by_t\bff_{t-k}\quad \text{and}\quad \wh\bSigma_{f}(k)=\frac{1}{T}\sum_{t=k+1}^T\bff_t\bff_{t-k}'\]
are the sample versions of $\bSigma_{yf}(k)$ and $\bSigma_{f}(k)$, respectively. For each $i$, we observe that there are $K+1$ unknown coefficients in the optimization problem (\ref{yw:est}), but there are only $K$ equations for each lag $k$ in (\ref{yw:est}), which implies that the optimization problem is not well-defined if we only make use of a single $k$ in the Yule-Walker estimation. To see this, we cast problem (\ref{yw:est})  into the framework of the generalized method of moments (GMM) (\cite{hansen1982large}). Let $\bff_{t-k}$ be the instrument, the moment conditions for (\ref{ft:sp}) are
\[E\bh_{t,k}(\rho_i,\bb_i)=0,\,\,\text{where}\,\,\bh_{t,k}(\rho_i,\bb_i)=(y_{i,t}-\rho_i\bw_i'\by_t-\bb_i'\bff_t)\bff_{t-k}',\,\,i=1,...,N,\]
which is equivalent to that in (\ref{pop:YW}) for each $k$. When $k=0$, it is not hard to see that 
\[\bh_{t,0}(\rho_i,\bb_i)=\frac{\partial \ve_{i,t}(\rho_i,\bb_i)}{\partial \bb_i},\]
where $\ve_{i,t}(\rho_i,\bb_i)=(y_{i,t}-\rho_i\bw_i'\by_t-\bb_i'\bff_t)^2$. Therefore, the equations produced by only taking partial derivatives concerning parameter $\bb_i$ are not sufficient to estimate $\rho_i$ and $\bb_i$ simultaneously. Then we conclude that estimation equations in (\ref{yw:est}) are not sufficient for any given $k\geq 0$.

To address this challenge, we combine two sets of estimating equations, resulting in $2K$ equations, which exceed the number of parameters by $K+1$ when $K \geq 1$. We first note the importance of cross-sectional dependence in asset returns and economic data, so we retain the $k = 0$ equations, which align with GMM using $\bff_t$ as instruments and the Least-Squares method. Additionally, since short-term dependence is more significant than long-term in economic and financial data, we focus on Equation (\ref{yw:est}) with $k = 0$ and $k = 1$. These lags capture the key dynamic information, reflecting the most relevant dependencies while excluding less impactful higher lags.

Specifically, let
\[\wh\bY_i=\left(\begin{array}{c}
     \wh\bSigma_{yf}'\be_i  \\
     \wh\bSigma_{yf}'(1)\be_i 
\end{array}\right)\,\,\text{and}\,\,\wh\bX_i=\left(\begin{array}{cc}
    \wh\bSigma_{yf}'\bw_i & \wh\bSigma_f \\
    \wh\bSigma_{yf}'(1)\bw_i & \wh\bSigma_f'(1)
\end{array}\right), \]
where $\wh\bSigma_{yf} = \wh\bSigma_{yf}(0)$ and $\wh\bSigma_f = \wh\bSigma_f(0)$. Due to the spatial nature of Model (\ref{ft:sp}), $\wh\bX_i$ is asymptotically singular, though not in finite samples. To address this, we apply ridge regression. Define $\bbeta = (\wh\rho, \bb')' \in \mathbb{R}^{K+1}$ and solve the following optimization problem for a given $\lambda_i > 0$:
\begin{equation}\label{YW:lse}
    \wh\bbeta_i(\lambda_{i})=(\wh\rho_i,\wh\bb_i')'=\arg\min_{\rho\in R, \bb\in R^K}\{\|\wh\bY_i-\wh\bX_i\bbeta\|_2^2+\lambda_{i}\|\bbeta\|_2^2\},\,\,i=1,...,N.
\end{equation}
The estimator has the explicit form:
\begin{equation}\label{lse:ob}
    \wh\bbeta_i(\lambda_i)=(\wh\bX_i'\wh\bX_i+\lambda_{i}\bI_{K+1})^{-1}\wh\bX_i'\wh\bY_i,\,\,i=1,...,N,
\end{equation}
which is the ridge estimator. In the subsequent analysis, we denote $(\wh\bX_i'\wh\bX_i)^+$ as the Moore-Penrose generalized inverse and let $\wh\bbeta_i = \wh\bbeta_i(0)$ as $\lambda_i \to 0$. Theorem~\ref{thm2} establishes the joint asymptotic distribution under these conditions, enabling joint inference. In finite samples, the estimator depends on the number of lagged auto-covariances used in the Yule-Walker estimation in (\ref{yw:est}), but its asymptotic convergence remains valid, as demonstrated in Section~\ref{sec3}.

\subsection{Boosting the Strength of Factor Instruments}\label{bst}

In practice, we stack only the cases when $k=0$ and $k=1$ as discussed in Section~\ref{sec22}. This 
approach provides a jointly consistent estimator and avoids unnecessary errors in the generalized method of moments estimation. The choice of $k=1$ is based on the assumption that short-term dependence is often 
more relevant than long-term dependence. If necessary, we can define a measure to select the optimal lag $k^*$ as follows:
\[k^*=\arg\max_{1\leq k\leq \bar{k}}|\det(\wh\bSigma_{f}(k))|,\]
where $\bar{k}$ is a small positive integer, and $|\det(\wh\bSigma_{f}(k))|$ is the product of the singular values of $\wh\bSigma_{f}(k)$. This measure captures the correlation strength between the lagged instruments and the contemporaneous factors. We then define
\[\wh\bY_{i,*}=\left(\begin{array}{c}
     \wh\bSigma_{yf}'\be_i  \\
     \wh\bSigma_{yf}'(k^*)\be_i 
\end{array}\right)\,\,\text{and}\,\,\wh\bX_{i,*}=\left(\begin{array}{cc}
    \wh\bSigma_{yf}'\bw_i & \wh\bSigma_f \\
    \wh\bSigma_{yf}'(k^*)\bw_i & \wh\bSigma_f'(k^*)
\end{array}\right), \]
and substitute them into (\ref{YW:lse}), yielding the refined estimator
\begin{equation}\label{beta:st}
    \wh\bbeta_{i,*}(\lambda_i)=(\wh\bX_{i,*}'\wh\bX_{i,*}+\lambda_{i}\bI_{K+1})^{-1}\wh\bX_{i,*}'\wh\bY_{i,*}, \quad i=1,...,N.
\end{equation}

\subsection{Estimation When Factors Are Latent}\label{sec23}	
		
In this section, we address the case where the factor processes $\bff_t$ are unobservable. We focus on latent factors that represent the internal dynamics driving the data $\by_t$, 
because the case with factors arising from some external data sources is similar to the diffusion-index framework in \cite{Stock2002} and \cite{gao2024supervised}. The estimation of the proposed model becomes more complex because, in addition to estimating the parameters in $\boldsymbol{\rho}$ and $\bB$, we must also recover the unknown factors. Note that
\begin{equation}\label{ft:latent}
    \by_t=(\bI_N-\bD(\boldsymbol{\rho})\bW)^{-1}\bB\bff_t+(\bI_N-\bD(\boldsymbol{\rho})\bW)^{-1}\bve_t=\bLambda\bff_t+\bxi_t,
\end{equation}
where $\bLambda=(\bI_N-\bD(\boldsymbol{\rho})\bW)^{-1}\bB$	and $\bxi_t=(\bI_N-\bD(\boldsymbol{\rho})\bW)^{-1}\bve_t$ are the loading matrix associated with the factor process $\bff_t$ and the idiosyncratic term, respectively.	Symbolically,  (\ref{ft:latent}) is a factor model with unknown factors and loading matrix, both of which need to be estimated from the data $\by_t$, for $t=1,...,T$.

Under the framework of Model (\ref{ft:latent}), we have a factor model with static factors 
and could use either the PCA method of \cite{bai2002determining} to estimate cross-sectional factors or the eigen-analysis method in \cite{lam2012factor} to extract dynamically dependent factors. However, the PCA method is not suitable for the spatial interactions in Model (\ref{ft:latent}), as the idiosyncratic noise recovered by PCA is often serially correlated. In contrast, the noise term $\bve_t$ (or $\bxi_t$) in the spatial model is white noise, which contradicts the PCA framework.


For spatial panel dynamic models in econometrics, the noise term $\bve_t$ (or $\bxi_t$) is white with zero serial correlation, while the dynamically dependent factors $\bff_t$ capture all the dynamic information of the data $\by_t$. This framework aligns with \cite{lam2011estimation}, \cite{lam2012factor}, and \cite{gao2022modeling}, among others. Based on the auto-covariance-based eigenanalysis in \cite{lam2011estimation}, we propose a two-step procedure to estimate the factors and other unknown coefficients, assuming the number of factors $K$ is known. The method for determining $K$ will be discussed later.

Note that $\bLambda$ and $\bff_t$ are not uniquely determined in (\ref{ft:latent}) and they require certain identification conditions. For simplicity, we assume that $\bLambda$ is a semi-orthogonal matrix scaled by $\sqrt{N}$ such that $\bLambda'\bLambda/N=\bI_K$. However, the loading and factors are still not uniquely identified because we can replace $(\bLambda,\bff_t)$ with $(\bLambda\bH, \bH'\bff_t)$ for any orthonormal matrix $\bH \in \mathbb{R}^{K\times K}$. Nevertheless, the linear space spanned by the columns of $\bLambda$, denoted $\mathcal{M}(\bLambda)$, is uniquely defined and referred to as the factor loading space.

Under the assumption that $\bve_t$ is a white noise process and $\Cov(\bff_t, \bve_{t+j}) = 0$, for $j \geq 0$, we allow for the possibility that $\bff_t$ may depend on the past lagged noises $\bve_{t-k}$, for some $k \geq 1$, as $\bff_t$ is a dynamically dependent process. For any integer $k \geq 1$, define the following covariance matrices of interest:
\[\bSigma_y(k)=\Cov(\by_t,\by_{t-k}),\,\,\bSigma_f(k)=\Cov(\bff_t,\bff_{t-k}),\,\,\text{and}\,\,\bSigma_{f\xi}(k)=\Cov(\bff_t,\bxi_{t-k}).\]
From (\ref{ft:latent}), we have 
\begin{equation}\label{sigy-r}
\bSigma_y(k)=\bLambda\bSigma_f(k)\bLambda'+\bLambda\bSigma_{f\xi}(k), \quad k\geq 1.
\end{equation}
For a pre-specified integer $k_0>0$, define
\begin{equation}\label{M}
\bM=\sum_{k=1}^{k_0}\bSigma_y(k)\bSigma_y'(k)=\bLambda\sum_{k=1}^{k_0}[\bSigma_{f}(k)\bLambda'+\bSigma_{f\xi}(k)][\bLambda\bSigma_f'(k)+\bSigma_{f\xi}'(k)]\bLambda',
\end{equation}
which is an $N \times N$ semi-positive definite matrix. Let $\bLambda_c$ denote the orthogonal complement matrix of $\bLambda$. We observe that $\bM \bLambda_c = \mathbf{0}$, implying that the columns of $\bLambda_c$ are the eigenvectors corresponding to the zero eigenvalues of $\bM$. The factor loading space $\mathcal{M}(\bLambda)$ is thus spanned by the eigenvectors (scaled by $\sqrt{N}$) corresponding to the $K$ non-zero eigenvalues of $\bM$. The integer $k_0$ in (\ref{M}) is a prescribed value that allows us to accumulate dynamic information across different lags. Since the dynamic dependence between $\by_t$ and $\by_{t-k}$ typically decreases as $k$ increases for stationary processes, a small $k_0$ is generally sufficient in practice. For further details on the rationale for using (\ref{M}) to estimate the loading space from a projection perspective, we refer readers to \cite{gao2021two}.


In practice, given the sample data $\{\by_t \mid t = 0, 1, \dots, T\}$, the first step of the procedure is to estimate the loading matrix $\bLambda$ or its column space $\mathcal{M}(\bLambda)$, and to recover the factor process $\bff_t$, assuming that the number of factors $K$ is known. The estimation of $K$ will be discussed later. Let $\wh\bSigma_y(k)$ denote the lag-$k$ sample autocovariance matrix of $\by_t$, defined similarly to those in (\ref{yw:est}). To estimate $\mathcal{M}(\bLambda)$, we perform an eigen-analysis of the sample version of $\bM$, defined as
\begin{equation}\label{mhat}
\wh\bM=\sum_{k=1}^{k_0}\wh\bSigma_y(k)\wh\bSigma_y'(k).
\end{equation}
Let $\wh\bLambda$ be the standardized semi-orthogonal matrix consisting of the eigenvectors of $\wh\bM$, scaled by $\sqrt{N}$, as its columns. The recovered factor processes are denoted as $\wh\bff_t = \frac{1}{N} \wh\bLambda' \by_t$, which can be obtained by the Ordinary Least Squares (OLS) method.

In the second step, we estimate the scalar coefficient vector $\boldsymbol{\rho}$ and the loading matrix $\bB$ in Model (\ref{ft:sp}). Let ${\wh\bff_1, \dots, \wh\bff_T}$ denote the estimated factors obtained in the first step. Define the following quantities:
  \[\wt\bSigma_{yf}(k)=\frac{1}{T}\sum_{t=k+1}^T\by_t\wh\bff_{t-k}',\,\,\wt\bSigma_f(k)=\sum_{t=k+1}^T\wh\bff_t\wh\bff_{t-k}',\,\,\text{and}\,\,\wt\bSigma_{\ve f}(k)=\frac{1}{T}\sum_{t=k+1}^T\bve_t\wh\bff_{t-k}',\]
and $\wt\bSigma_{yf}=\wt\bSigma_{yf}(0)$, $\wt\bSigma_f=\wt\bSigma_f(0)$, and $\wt\bSigma_{\ve f}=\wt\bSigma_{\ve f}(0)$. Following a similar procedure to the shrinkage Yule-Walker estimation in Section~\ref{sec22}, where the factors are observable, we formulate the following optimization problem for the case of augmenting only $k=0$ and $k=1$, with a given $\lambda_i > 0$:
	\begin{equation}\label{yw:latent}
    \wt\bbeta_i(\lambda_i)=(\wt\rho_i,\wt\bb_i')'=\arg\min_{\rho\in R, \bb\in R^r}\{\|\wt\bY_i-\wt\bX_i\bbeta\|_2^2+\lambda_{i}\|\bbeta\|_2^2\},\,\,i=1,...,N,
\end{equation}
where 
\[\wt\bY_i=\left(\begin{array}{c}
     \wt\bSigma_{yf}'\be_i  \\
     \wt\bSigma_{yf}'(1)\be_i 
\end{array}\right)\,\,\text{and}\,\,\wt\bX_i=\left(\begin{array}{cc}
    \wt\bSigma_{yf}'\bw_i & \wt\bSigma_f' \\
    \wt\bSigma_{yf}'(1)\bw_i & \wt\bSigma_f'(1)
\end{array}\right) \]
represent the response variables and covariates, respectively.
The Yule-Walker estimation in (\ref{yw:latent}) then yields the least squares (LS) estimator for $\bbeta$ as
\begin{equation}\label{lse:lat}
 \wt\bbeta_i(\lambda_i)=(\wt\bX_i'\wt\bX_i+\lambda_{i}\bI_{K+1})^{-1}\wt\bX_i'\wt\bY_i,\,\,i=1,...,N.
\end{equation}
Thus, we perform $N$ Yule-Walker estimation procedures, for $i = 1, \dots, N$ and obtain the estimators $\tilde{\boldsymbol{\rho}} = (\wt\rho_1, \dots, \wt\rho_N)'$ and $\wt\bB = (\wt\bb_1, \dots, \wt\bb_N)'$. We can similarly define $\wt\bbeta_i = \wt\beta_i(0)$ by adopting the Moore-Penrose inverse of $\wt\bX_i' \wt\bX_i$ as in (\ref{lse:ob}). Theorem~\ref{thm5} in Section~\ref{sec3} establishes the joint asymptotic distribution, which can be utilized for joint inference under this condition.

In practice, we may also use the boosting method described in Section~\ref{bst} to select the optimal lag $k*$. We can then replace $\wt\bSigma_{yf}(1)$ and $\wt\bSigma_f(1)$ with $\wt\bSigma_{yf}(k^*)$ and $\wt\bSigma_f(k^*)$, respectively, in $\wt\bY_i$ and $\wt\bX_i$. The estimator  $\wt\bbeta_{i,*}(\lambda_i)$ can be obtained in the same manner as that described in Section~\ref{bst}.

  \subsection{Selecting the Number of Factors and the Penalty Parameters}
  
In this section, we discuss the determination of the number of factors 
$K$ in Model (\ref{ft:latent}), which is typically unknown in practice. Over the past decades, several methods have been developed to estimate 
$K$, including the information criteria proposed by \cite{bai2002determining}, the random matrix theory approach in \cite{onatski2010determining}, the ratio-based method in \cite{lam2012factor} and \cite{Ahn2013}, the canonical correlation analysis technique in \cite{gao2019structural}, and the white noise testing approach in \cite{gao2022modeling}, among others. 
In this paper, we introduce two widely used methods for estimating $K$.

The first method is an information criterion introduced by \cite{bai2002determining}. It estimates $K$ by
\begin{equation}\label{ri}
\wh K=\arg\min_{0\leq j\leq J}\log(\frac{1}{NT}\sum_{t=1}^T\|\by_t-\frac{1}{N}\wh\bLambda_{j}\wh\bLambda_j'\by_t\|_2^2)+jg(T,N),
\end{equation}
where $J$ is a prescribed upper bound, $\wh\bLambda_{j}$ is a $N\times j$ estimated loading matrix, and $g(T,N)$ is a penalty function of $(N,T)$ such that $g(T,N)=o(1)$ and $\min\{N,T\}g(T,N)\rightarrow\infty$. Two examples of $g(T,N)$ suggested by \cite{bai2002determining} are IC1 and IC2 given below:
\[IC1=\frac{N+T}{NT}\log(\frac{NT}{N+T})\quad \text{and}\quad IC2=\frac{N+T}{NT}\log(\min\{N,T\}).\]

For the estimation of $K$, in addition to the information criterion in (\ref{ri}), we can adopt the ratio-based method proposed in \cite{lam2012factor} and \cite{Ahn2013}. Let $ \wh\mu_1 \geq \dots \geq \wh\mu_{N} $ be the $N$ eigenvalues of $\wh\bM$. We estimate $K$ by
\begin{equation}\label{est-r} \wh K=\arg\min_{1\leq l\leq R}{{\wh\mu_{l+1}}/{\wh\mu_{l}}}, \end{equation}
where $R = \lfloor N/2 \rfloor$ is commonly used, as suggested by \cite{lam2012factor}.

For the selection of the penalty parameter $\lambda_i$, it is common to assume that $\lambda_i \in \mathcal{S}$, where $\mathcal{S}$ is a candidate set consisting of possible penalty choices. We split the data sample into two segments, ${\by_1, \dots, \by_{T_1}}$ and ${\by_{T_1+1}, \dots, \by_{T}}$. Suppose $\wh\rho_i(\lambda)$ and $\wh\bb_i(\lambda)$ are the estimators obtained from the first segment. The optimal $\lambda$ is chosen by solving
\begin{equation}
    \wh\lambda_i=\argmin_{\lambda\in\mathcal{S}}\frac{1}{T-T_1}\sum_{t=T_1+1}^{T}\|y_{i,t}-\wh\rho_i(\lambda)\bw_i'\by_{t}-\wh\bb_i(\lambda)\bff_t\|_2^2.
    \end{equation}
When the factors are unobservable, we replace $\bff_t$ with $\wh\bff_t$, which is estimated from the second segment using the estimator $\wh\bLambda(\lambda)$ obtained from the first segment.

	\section{Theoretical Properties}\label{sec3}	

In this section, we present the asymptotic theory for the estimation method of  Section~\ref{sec2}, when both the dimension $N$ and the sample size $T$ tend to infinity. We focus on the estimating equations with lags $k = 0$ and $k = 1$, which typically capture the majority of the cross-sectional and dynamic dependencies in the data. A  constant $C$ is used generically, with its value potentially varying across different parts of the analysis. We begin with some assumptions.
    

	
\begin{assumption}\label{asm1}
The process $\{(\by_t,\bff_t)\}$ is strictly stationary and $\alpha$-mixing with the mixing coefficient satisfying the condition $\sum_{k=1}^\infty\alpha_N(k)^{1-2/\gamma}<\infty$ for some $\gamma>2$, where
\[\alpha_N(k)=\sup_{i}\sup_{A\in\mathcal{F}_{-\infty}^i,B\in \mathcal{F}_{i+k}^\infty}|P(A\cap B)-P(A)P(B)|,
\]
and $\mathcal{F}_i^j$ is the $\sigma$-field generated by $\{(\by_t,\bff_t):i\leq t\leq j\}$.
\end{assumption}
\begin{assumption}\label{asm2}
    The spatial weight matrix $\bW$ is known with zero main diagonal elements, and the matrix $\bS_N(\boldsymbol{\rho}):=\bI_N-\bD(\boldsymbol{\rho})\bW$ is invertible. The row and column sums of $|\bW|$ and $|\bS_N(\boldsymbol{\rho})^{-1}|$ are bounded uniformly in $N$.
\end{assumption}
\begin{assumption}\label{asm3}
    $\{\bve_t\}$ is a white noise process satisfying $\Cov(\by_{t-j},\bve_t)={\bf 0}$ and $\Cov(\bff_{t-k},\bve_t)={\bf 0}$, for $j\geq 1$ and $k\geq 0$, respectively.
\end{assumption}
\begin{assumption}\label{asm4}
    (i) If $\bff_t$'s are observed, each element in $\bB$ are bounded uniformly in $N$; (ii) If  $\bff_t$'s are latent, the loading matrix $\bB$ is of full rank 
    such that $\frac{1}{N}\bB'\bS_N'(\boldsymbol{\rho})^{-1}\bS_N(\boldsymbol{\rho})^{-1}\bB=\bI_r$, which is an identity matrix.
\end{assumption}
\begin{assumption}\label{asm5}
    For  $1\leq j\leq K$ and $1\leq k\leq N$, $E|f_{j,t}|^{2\gamma}<C$ and $E|\ve_{k,t}|^{2\gamma}<C$, where $\gamma$ is given in Assumption~\ref{asm1}.
\end{assumption}

\begin{assumption}\label{asm6}
    For $i=1,...,N$, the rank of matrix $\bX_i'
    \bX_i+\lambda\bI_{K+1}$ is $K+1$, for any $\lambda>0$, where
    \[\bX_i=\left(\begin{array}{cc}
    \bSigma_{yf}'\bw_i & \bSigma_f' \\
    \bSigma_{yf}'(1)\bw_i & \bSigma_f'(1)
\end{array}\right). \]
\end{assumption}
Assumption \ref{asm1} is standard for dependent random processes. See \cite{gao2019banded} for a theoretical justification for VAR models. In fact, the assumption of strict stationarity can be removed and we only need to replace definitions of $\bSigma_y(k)$ and $\bSigma_f(k)$ with $\frac{1}{T}\sum_{t=k+1}^T\Cov(\by_t,\bff_{t-k})$ and $\frac{1}{T}\sum_{t=k+1}^T\Cov(\bff_t,\bff_{t-k})$, respectively, and the results still hold throughout the paper. Assumption \ref{asm2} is commonly used in the spatial econometrics literature to limit the dependence across different locations or economic units; see, for example, \cite{lee2010some}. Assumption~\ref{asm3} is weaker than the independence assumptions imposed in the spatial econometrics literature and we also allow for possible dependence between $\by_{t+j}$ and $\bff_{t+k}$ and past lagged of noises, for $j\geq 0$ and $k\geq 1$.
Assumption~\ref{asm4} is standard for the loading matrix under the scenarios when the factors are either observed  or latent. Assumption~\ref{asm5} imposes some moment conditions on the factors and noise terms. It is not hard to see that $E|y_{i,t}|^{2\gamma}<C$ under Assumptions~\ref{asm2}, \ref{asm4} 
and \ref{asm5}. Furthermore, this also implies that {$E|\bw_i'\bSigma_{yf}\bff_t|^{2\gamma} < C$, $E|\bw_i'\bSigma_{yf}(1)\bff_{t-1}|^{2\gamma} < C$, $E\|\bSigma_f\bff_t\|_2^{2\gamma} < C$, and $E\|\bSigma_f(1)\bff_{t-1}\|_2^{2\gamma} < C$}, which are used to establish the convergence of the variance of $\bS_{N,T}$, as defined in (\ref{snt}) of the online Appendix. Assumption~\ref{asm6} ensures that the ridge solutions in (\ref{lse:ob}) and (\ref{lse:lat}) are well-defined. 


Now, we present the asymptotic properties of $\wh\bbeta_i$, for $i = 1, \dots, N$. 

	\begin{theorem}\label{thm1}
	    Let Assumptions \ref{asm1} $-$ \ref{asm6} hold. \\
     (i) If $N=o(T)$, we have
     \[\|\wh\bbeta_i(\lambda_i)-\wh\bX_i(\lambda_i)^{-1}\wh\bX_i\wh\bX_i'\bbeta_i\|_2=O_p(T^{-1/2}),\quad i=1,...,N,\]
     as $N,T\rightarrow\infty$, where $\wh\bX_i(\lambda_i)=\wh\bX_i'\wh\bX_i+\lambda_i\bI_{K+1}$.\\
     (ii) If $N=o(T)$ and let $\lambda_i\rightarrow 0$, we have
     \[\|(\wh\bX_i'\wh\bX_i)(\wh\bbeta_i-\bbeta_i)\|_2=O_p(T^{-1/2}),\quad i=1,...,N,\]
     as $N,T\rightarrow\infty$.
	\end{theorem}	
Theorem~\ref{thm1} implies that the ridge estimator for $\bbeta$ is biased, which is a common issue in ridge estimation. However, we can establish the joint convergence of $\wh\rho_i$ and $\bb_i$, as stated in Theorem~\ref{thm1}(ii). Since $\rho_i$ and $\bb_i$ represent loadings for all possible factors, this result is useful because these coefficients can be jointly estimated and inferred in many economic contexts, such as financial networks, as described in \cite{wang2021joint}.

Next, we provide the joint limiting distributions of the shrinkage estimators. For $i=1,...,N$, define
\[\bSigma_{f\ve_i}(0,0)=\Cov(\bff_t\ve_{i,t},\bff_t\ve_{i,t}),\,\,\bSigma_{f\ve_i}(1,0)=\Cov(\bff_t\ve_{i,t},\bff_{t-1}\ve_{i,t}),\,\,\bOmega_{f\ve_i}(0,0)=\Cov(\bff_{t-1}\ve_{i,t},\bff_{t-1}\ve_{i,t}),\]
\[\bSigma_{f\ve_i}(k,j)=\Cov(\bff_{t+j}\ve_{i,t+j},\bff_{t-k}\ve_{i,t})+\Cov(\bff_t\ve_{i,t},\bff_{t-k+j}\ve_{i,t+j}),\,\,j\geq 1,k\geq 0,\]
\[\bOmega_{f\ve_i}(0,j)=\Cov(\bff_{t-1+j}\ve_{i,t+j},\bff_{t-1}\ve_{i,t})+\Cov(\bff_{t-1}\ve_{i,t},\bff_{t-1+j}\ve_{i,t+j}),\quad j\geq 1,\]
\[\bSigma_{f\ve_i}(0)=\sum_{j=0}^{\infty}\bSigma_{f\ve_i}(0,j), \,\,\bSigma_{f\ve_i}(1)=\sum_{j=0}^{\infty}\bSigma_{f\ve_i}(1,j), \,\,\bOmega_{f\ve_i}(0)=\sum_{j=0}^{\infty}\bOmega_{f\ve_i}(0,j).\]
Let
\begin{equation}\label{Vi}
 \bV_i=\left(\begin{array}{cc}
     \bw_i'\bSigma_{yf}\bSigma_{yf}'\bw_i+\bw_i'\bSigma_{yf}(1)\bSigma_{yf}'(1)\bw_i &\bw_i'\bSigma_{yf}\bSigma_f+\bw_i'\bSigma_{yf}(1)\bSigma_f'(1) \\
  \bSigma_f\bSigma_{yf}'\bw_i+\bSigma_f(1)\bSigma_{yf}'(1)\bw_i   &\bSigma_f^2+\bSigma_f(1)\bSigma_f'(1) 
\end{array}\right)  
\end{equation}
and
\begin{equation}\label{Ui}
 \bU_i=\left(\begin{array}{cc}
     \bSigma_{f\ve_i}(0) &\bSigma_{f\ve_i}(1) \\
  \bSigma_{f\ve_i}'(1)  &\bOmega_{f\ve_i}(0)
\end{array}\right).
\end{equation}

The following theorem establishes the joint asymptotic normality of the estimators.
\begin{theorem}\label{thm2}
     Let Assumptions \ref{asm1} $-$ \ref{asm6} hold.  If $N=o(T)$ and $\lambda_i\rightarrow 0$, we have
     \[\sqrt{T}\bV_i(\wh\bbeta_i-\bbeta_i)\longrightarrow_d N({\bf 0}, \bX_i'\bU_i\bX_i),\]
     for $i=1,...,N$ as $N,T\rightarrow\infty$, where $\bU_i$ and $\bV_i$ are defined in (\ref{Ui}) and (\ref{Vi}), respectively.
 \end{theorem}
From Theorem~\ref{thm2}, we see that the Yule-Walker estimators obtained in (\ref{lse:ob}) are asymptotically normal when the dimension $N$ diverges. The convergence rate is the standard $\sqrt{T}$ under the assumption that $N/T \rightarrow 0$, which is a similar requirement in spatial panel dynamic models; see \cite{yu2008quasi}, among others. The condition $N/T \rightarrow 0$ is weaker than the one in \cite{gao2019banded}, where $N/\sqrt{T} \rightarrow 0$ is required, because we assume the dimension of $\bff_t$ is $K$, a finite integer. The convergence of $\wh\bSigma_{yf}(k)$ to $\bSigma_{yf}(k)$ only requires $N/T \rightarrow 0$, whereas the convergence of $\wh\bSigma_y(k) = \frac{1}{T}\sum_{t=k+1}^T \by_t \by_{t-k}'$ to its population version requires $N/\sqrt{T} \rightarrow 0$ as stated in \cite{gao2019banded}.
By the form of $\bX_i$ in Assumption~\ref{asm6}, we can show that
\begin{equation}\label{xux}
\bX_i'\bU_i\bX_i=\left(\begin{array}{cc}
     \bSigma_{i,11} &\bSigma_{i,12} \\
  \bSigma_{i,21}  &\bSigma_{i,22}
\end{array}\right), 
\end{equation}
where
\begin{align*}
\bSigma_{i,11}=&\bw_i'\bSigma_{yf}\bSigma_{f\ve_i}(0)\bSigma_{yf}'\bw_i+\bw_i'\bSigma_{yf}\bSigma_{f\ve_i}(1)\bSigma_{yf}'(1)\bw_i+\bw_i'\bSigma_{yf}(1)\bSigma_{f\ve_i}'(1)\bSigma_{yf}'\bw_i\notag\\
&+ \bw_i'\bSigma_{yf}(1)\bOmega_{f\ve_i}(0)\bSigma_{yf}'(1)\bw_i,
\end{align*}
\begin{align*}
\bSigma_{i,22}=&\bSigma_f\bSigma_{f\ve_i}(0)\bSigma_f+\bSigma_f\bSigma_{f\ve_i}(1)\bSigma_f'(1)+\bSigma_f(1)\bSigma_{f\ve_i}'(1)\bSigma_f+ \bSigma_f(1)\Omega_{f\ve_i}(0)\bSigma_f'(1),
\end{align*}
\begin{align*}
\bSigma_{i,12}=&\bw_i'\bSigma_{yf}\bSigma_{f\ve_i}(0)\bSigma_f+\bw_i'\bSigma_{yf}\bSigma_{f\ve_i}(1)\bSigma_f'(1)+\bw_i'\bSigma_{yf}(1)\bSigma_{f\ve_i}'(1)\bSigma_{f}\notag\\
&+ \bw_i'\bSigma_{yf}(1)\bOmega_{f\ve_i}(0)\bSigma_{f}'(1),
\end{align*}
and $\bSigma_{i,21}=\bSigma_{i,12}'$. These matries can all be estimated from the data.

 Finally, we turn to the case when the factors are latent. We need to make two more assumptions to establish the uniform convergence and the limiting distributions of the estimated factors.
 \begin{assumption}\label{asm7}
     $\bff_t$ and $\bve_t$ are sub-exponentially distributed in the sense that 
     \[P(|\bv_1'\bff_t|>x)\leq C\exp(-Cx),\,\,\text{and}\,\, P(|\bv_2'\bve_t|>x)\leq C\exp(-Cx),\]
     for any $x>0$, where $\|\bv_1\|_2=1$ and $\|\bv_2\|_2=1$ are any two constant vectors.
 \end{assumption}
 \begin{assumption}\label{asm8}
For each $t=1,...,T$, as $N\rightarrow\infty$,
\[\frac{1}{\sqrt{N}}\sum_{i=1}^N\bp_i\ve_{i,t}\longrightarrow_d N(0,\bGamma_t),\]
where $\bp_i$ is the $i$th column of $\bLambda'(\bI_N-\bD(\boldsymbol{\rho})\bW)^{-1}$, and $\bGamma_t=\lim_{N\rightarrow\infty}\frac{1}{N}\sum_{i=1}^N\sum_{j=1}^N\bp_i\bp_j'E(\ve_{i,t}\ve_{j,t})$ in probability.
 \end{assumption}
Assumption~\ref{asm7} is commonly used in the statistical and econometrics literature to establish uniform convergence. The sub-exponential distribution is a broader class of distributions than the sub-Gaussian distribution and includes the uniform distribution over every convex body, following the Brunn-Minkowski inequality. For further details, see, for example, \cite{vershynin2018high}. Assumption~\ref{asm8} is similar to Assumption F(3) in \cite{bai2003inferential}, which is used to establish the limiting distribution of the estimated factors.

We first state the convergence of the estimated loading matrix below, where we introduce a rotational matrix $\bH_{NT}$ in the proof of the following theorem. This approach differs from the techniques used in \cite{lam2011estimation}.

\begin{theorem}\label{thm3}
    Let Assumptions \ref{asm1} $-$ \ref{asm6} hold. If $N=o(T)$, then there exists an invertible matrix $\bH_{NT}$ such that
    \[\frac{1}{\sqrt{N}}\|\wh\bLambda-\bLambda\bH_{NT}'\|_F=O_p(\frac{1}{\sqrt{T}}).\]
\end{theorem}
\begin{remark}
    (i) Unlike the proof in \cite{lam2011estimation} where a matrix perturbation theory is used to show the convergence of the estimated loading matrix, we developed a new approach in the Appendix to show the convergence rate of $\wh\bLambda$. One of the advantages of the new approach is that we can specify the rotational matrix $\bH_{NT}$ which is defined as
    \[\bH_{NT}'=\sum_{k=1}^{k_0}\bG_{1,k}\bG_{1,k}'\bLambda'\wh\bLambda\wh\bV_{NT}^{-1}, \,\text{where}\,\, \bG_{1,k}=\frac{1}{T}\sum_{t=k+1}^T(\bff_t\bff_{t-k}'\bLambda'+\bff_t\bxi_{t-k}'),\]
    and $\wh\bV_{NT}\in R^{r}$ is a diagonal matrix with diagonal elements being the top $K$ eigenvalues of $\wh\bM$. See the proof of Theorem~\ref{thm3} in the online Appendix for details.\\
    (ii) Note that we impose that $\bLambda'\bLambda / N = \bI_r$, whereas \cite{lam2011estimation} assumes that $\bLambda'\bLambda = \bI_r$. Therefore, the convergence rate is the same as the one in Theorem 1 of \cite{lam2011estimation}, where we assume $\delta = 0$ in our paper, corresponding to the case of strong factors.
\end{remark}

Next, we establish the uniform convergence of the estimated factors and the corresponding limiting distributions.
\begin{theorem}\label{thm4}
    Let Assumptions \ref{asm1} $-$ \ref{asm6} hold.\\
    (i) If $\bve_t$ and $\bff_t$ are sub-exponentially distributed as in Assumption~\ref{asm7}, then there exists an invertible matrix $\bK_{NT} \in \mathbb{R}^{r}$ such that
    \[\max_{1\leq t\leq T}\|\wh\bff_t-\bK_{NT}\bff_t\|_2=O_p\{(\frac{1}{\sqrt{N}}+\frac{1}{\sqrt{T}})\log(T)\}.\]
    (ii) Let Assumption~\ref{asm8} also hold. If $N = o(T)$, then there exists an invertible matrix $\bK_{NT} \in \mathbb{R}^r$ and its limit $\bH \in \mathbb{R}^r$ such that
    \[\sqrt{N}(\wh\bff_t-\bK_{NT}\bff_t)=\bH\frac{1}{\sqrt{N}}\sum_{i=1}^N\bp_i\ve_{i,t}+o_p(1)\longrightarrow_d N(0,\bH\bGamma_t\bH'),\]
    where $\bp_i$ is the $i$th column of $\bLambda'(\bI_N - \bD(\boldsymbol{\rho}) \bW)^{-1}$, $\bH$ is the limit of $\bH_{NT}$ as shown in Lemma 2 of the online Appendix, and $\bGamma_t$ is defined as in Assumption \ref{asm8}.
\end{theorem}
\begin{remark}
    (I) A remarkable feature in Theorem~\ref{thm4} is that we only require $N/T \to 0$, and the asymptotic normality of $\bff_t$ can still be achieved. \\
    (ii) Note that we adopt the matrix $\bK_{NT}$ as a rotational matrix for $\bff_t$, which is defined as
    \[\bK_{NT}=\frac{1}{N}\wh\bLambda'\bLambda.\]
    See the proof of Theorem~\ref{thm4} in the Appendix. In fact, according to Lemma 1 of the Appendix, we may replace $\bK_{NT}$ by $\bH_{NT}$, and the results in Theorem~\ref{thm4} still hold. This can be shown by rewriting the term $\frac{1}{N} \wh\bLambda' \bLambda \bff_t$ in (\ref{fhat}) as
    \[\frac{1}{N}(\wh\bLambda-\bLambda\bH_{NT}')'\bLambda\bff_t+\bH_{NT}\bff_t,\]
   where the first term is still asymptotically negligible. However, we do not adopt this formula since it will introduce a bias term in establishing the limiting distributions of the $\wh\bff_t$. Nevertheless, it is not hard to show that $\bK_{NT}$ and $\bH_{NT}$ have the same limit as $N, T \to \infty$.
\end{remark}

Furthermore, we study the limiting distributions of the estimated parameters in (\ref{lse:lat}). Similar to the case when the factors are observable, we provide some notation used in the following Theorem. Let
\begin{equation}\label{Vih}
  \bV_i^H=\left(\begin{array}{cc}
     \bw_i'\bSigma_{yf}\bSigma_{yf}'\bw_i+\bw_i'\bSigma_{yf}(1)\bSigma_{yf}'(1)\bw_i &\bw_i'\bSigma_{yf}\bSigma_f\bH'+\bw_i'\bSigma_{yf}(1)\bSigma_f'(1)\bH' \\
  \bH\bSigma_f\bSigma_{yf}'\bw_i+\bH\bSigma_f(1)\bSigma_{yf}'(1)\bw_i   &\bH\bSigma_f^2\bH'+\bH\bSigma_f(1)\bSigma_f'(1)\bH' 
\end{array}\right)
\end{equation}
and
\begin{equation}\label{Uih}
   \bU_i^H=\left(\begin{array}{cc}
     \bH\bSigma_{f\ve_i}(0)\bH' &\bH\bSigma_{f\ve_i}(1)\bH'\\
  \bH\bSigma_{f\ve_i}'(1)\bH' &\bH\bOmega_{f\ve_i}(0)\bH'
\end{array}\right).
\end{equation}
The following theorem establishes the asymptotic normality of the estimators in (\ref{lse:lat}) when the factors are latent and the dimension $N$ is diverging.
 \begin{theorem}\label{thm5}
     Let Assumptions \ref{asm1} $-$ \ref{asm8} hold. \\
     (i) If $N=o(T)$ and $\sqrt{T}=o(N)$, then there exists an invertible matrix $\bK_{NT}\in R^r$ such that
    \[\wt\bbeta_i(\lambda_i)-\wt\bX_i(\lambda_i)^{-1}\wt\bX_i'\wt\bX_i\bK_{NT}^{*}\bbeta_i=O_p(T^{-1/2}),\]
     where $\wt\bX_i(\lambda_i)=\wt\bX_i'\wt\bX_i+\lambda_i\bI_{K+1}$.\\
     (ii) If $N=o(T)$ and $\sqrt{T}=o(N)$, let $\lambda_i\rightarrow 0$, there exists an invertible matrix $\bK_{NT}\in R^r$ such that
     \[\sqrt{T}\bV_i^H(\wt\bbeta_i-\bK_{NT}^*\bbeta_i)\longrightarrow_d N({\bf 0}, \bX_i^H{'}\bU_i^H\bX_i^H),\]
     for $i=1,...,N$ as $T\rightarrow\infty$, where $\bK_{NT}^{*}=\diag(1,(\bK_{NT}')^{-1})$ is a block-diagonal matrix, and $\bU_i^H$ and $\bV_i^H$ are defined in (\ref{Uih}) and (\ref{Vih}), respectively, and
     \[\bX_i^H=\left(\begin{array}{cc}
    \bH\bSigma_{yf}'\bw_i & \bH\bSigma_f\bH' \\
    \bH\bSigma_{yf}'(1)\bw_i & \bH\bSigma_f'(1)\bH'
\end{array}\right).\]
 \end{theorem}
\begin{remark}
    (i) From Theorem~\ref{thm5}, we see that the convergence rate is still the standard $\sqrt{T}$, which is the same as that in Theorem~\ref{thm2} when the factors are observable. On the other hand, we note that the scalar coefficient can be uniquely determined, but the coefficient vector $\bb_i$ can be estimated up to a rotational matrix $\bK_{NT}$, which is reasonable due to the identification issue in the factor analysis.\\
    (ii) Recall that this is a two-step procedure. The statistical inference is usually difficult to establish in the second step because the errors incurred in the first step sometimes create a biased term. As discussed in Remark 2(ii), we adopt a rotational matrix $\bK_{NT}$ instead of $\bH_{NT}$ in Theorems~\ref{thm4} and \ref{thm5} such that the bias term can be erased, although $\bK_{NT}$ and $\bH_{NT}$ have the same limit. See the proof of Theorem~\ref{thm5} in the online Appendix for details.
\end{remark}
It can be easily shown that the variance term in Theorem~\ref{thm5}(ii) can be expressed as
\[\bX_i^{H}{'}\bU_i^H\bX_i^H=\left(\begin{array}{cc}
     \bSigma_{i,11}^H &\bSigma_{i,12}^H \\
  \bSigma_{i,21}^H  &\bSigma_{i,22}^H
\end{array}\right), \]
where
\begin{align*}
\bSigma_{i,11}^H=&\bw_i'\bSigma_{yf}\bSigma_{f\ve_i}(0)\bSigma_{yf}'\bw_i+\bw_i'\bSigma_{yf}\bSigma_{f\ve_i}(1)\bSigma_{yf}'(1)\bw_i+\bw_i'\bSigma_{yf}(1)\bSigma_{f\ve_i}'(1)\bSigma_{yf}'\bw_i\notag\\
&+ \bw_i'\bSigma_{yf}(1)\bOmega_{f\ve_i}(0)\bSigma_{yf}'(1)\bw_i,
\end{align*}
\begin{align*}
\bSigma_{i,22}^H=&\bH\bSigma_f\bSigma_{f\ve_i}(0)\bSigma_f\bH'+\bH\bSigma_f\bSigma_{f\ve_i}(1)\bSigma_f'(1)\bH'+\bH\bSigma_f(1)\bSigma_{f\ve_i}'(1)\bSigma_f\bH'+ \bH\bSigma_f(1)\Omega_{f\ve_i}(0)\bSigma_f'(1)\bH',
\end{align*}

\begin{align*}
\bSigma_{i,12}^H=&\bw_i'\bSigma_{yf}\bSigma_{f\ve_i}(0)\bSigma_f\bH'+\bw_i'\bSigma_{yf}\bSigma_{f\ve_i}(1)\bSigma_f'(1)\bH'+\bw_i'\bSigma_{yf}(1)\bSigma_{f\ve_i}'(1)\bSigma_{f}\bH'\notag\\
&+ \bw_i'\bSigma_{yf}(1)\bOmega_{f\ve_i}(0)\bSigma_{f}'(1)\bH',
\end{align*}
and $\bSigma_{i,21}^H=\bSigma_{i,12}^H{'}$.

The consistency of the estimated number of factors using the information criterion in (\ref{ri}) or the ratio-based method in (\ref{est-r}) can be established by a standard argument as that in \cite{bai2002determining} or \cite{Ahn2013}. We omit the details.

\begin{remark}
 In the estimation procedure above, we primarily focus on the augmented method by stacking factor lags for $k=0$ and $k=1$. In fact, ridge regression can be applied by taking any finite number of lagged factors in the Yule-Walker estimation. The theory can be established in a similar way.

\end{remark}

\begin{remark}
The QMLE method proposed in \cite{aquaro2021estimation} and \cite{hu2023arbitrage} can yield pointwise consistent estimators but is feasible only when the dimension $N$ is small and fixed. Additionally, they only focus on cases when the factors are observable. As $N$ increases, additional bias can arise, and the asymptotic results do not hold anymore (see Remarks 6–7 in \cite{aquaro2021estimation}). Moreover, the computational cost of the QMLE method becomes prohibitive for large $N$. In contrast, the proposed generalized Yule-Walker method is designed to handle scenarios with large or diverging $N$ while remaining computationally efficient. Simulations and real data analyses in Sections~\ref{sec4}-\ref{sec5} show that the proposed method can even outperform the QML approach, achieving smaller out-of-sample forecasting errors.
\end{remark}

 \section{Simulation Studies}\label{sec4}
In this section, we use Monte Carlo simulations to evaluate the performance of the proposed methodology across a spectrum of finite samples.

Consider the  model in Section \ref{sec2} with common factors generated from a VAR(1) process $\bff_t = \boldsymbol{\Phi} \bff_{t-1} +\boldsymbol{\eta}_t$. Here,  $\boldsymbol{\Phi}$ is a diagonal matrix, with entries independently sampled from a uniform distribution $ U (0.5, 0.9)$ and the error term  $\boldsymbol{\eta_t} \sim N (0, \bI_K)$.
For each  realization of $\by_t$, the elements of the loading matrix $\bB$ are independently drawn from $U (-2, 2)$, and the idiosyncratic error term $\bve_t $ is generated from $N (0, I_N)$. The spatial $\brho$ is sampled independently from a power-law distribution with an exponent $\alpha=5$.  To construct the spatial matrices, the $q$ neighboring off-diagonal elements are set to 1 and  the diagonal elements are 0, followed by row normalization to ensure each row sums to 1.  We set $q=3$  and the true number of factors $K = 3$, with dimension $N = 25,50,100,200$, and sample size $T = 50,100,200,400,1000$. We use 1000 replications for each configuration of $(T,N)$. To make the results below replicable, the seed is set to be \texttt{1234} in the \texttt{R} programming.

\begin{figure}[!ht]
  \includegraphics[width = \textwidth]{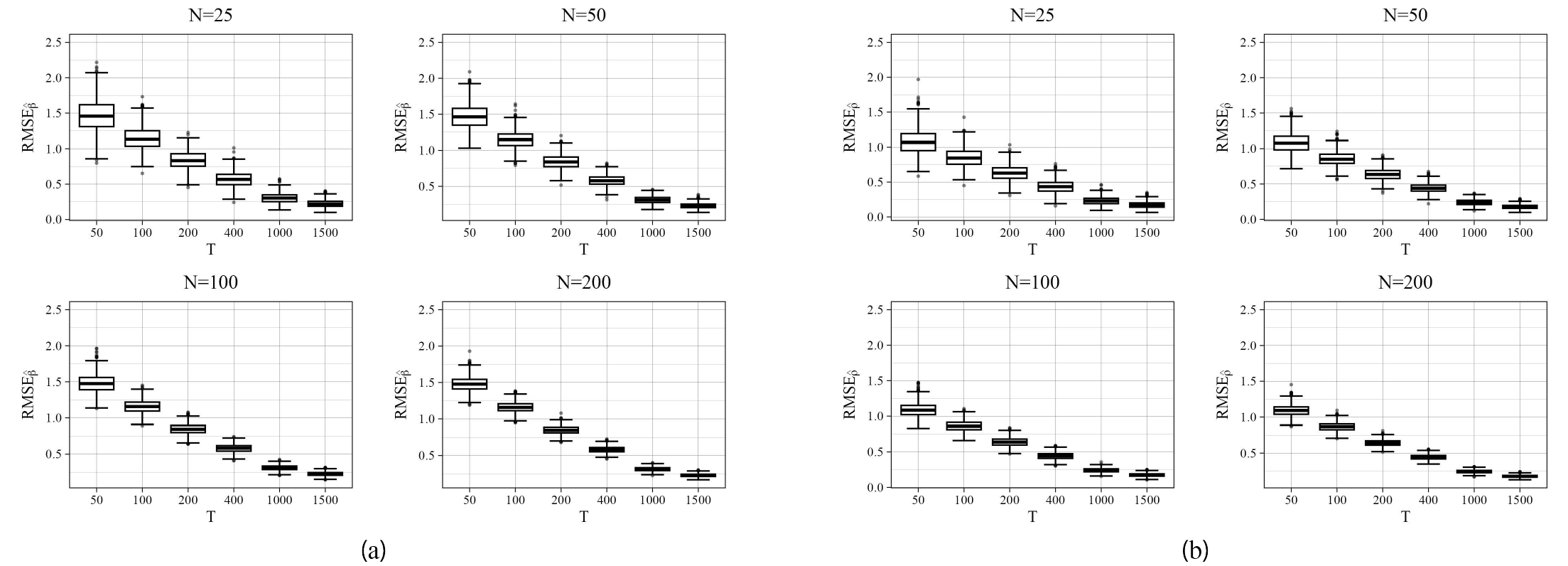}
  \caption{Boxplots of estimator convergence for Eq~(\ref{YW:lse}) with a fixed large ridge penalty parameter ($\lambda_i = 10^{-3}$), where $N$ and $T$ denote the dimension and sample size, 
  respectively. (a) shows the joint estimation performance of \( \widehat{\bbeta}_i \) measured by $\mathrm{RMSE}_{\hat{\beta}}$, and (b) shows the estimation performance of the spatial parameter $\wh\brho$ measured by $\mathrm{RMSE}_{\hat{\rho}}$.}\label{fig:thm2_1}
\end{figure}

\begin{figure}[ht]
  \includegraphics[width = \textwidth]{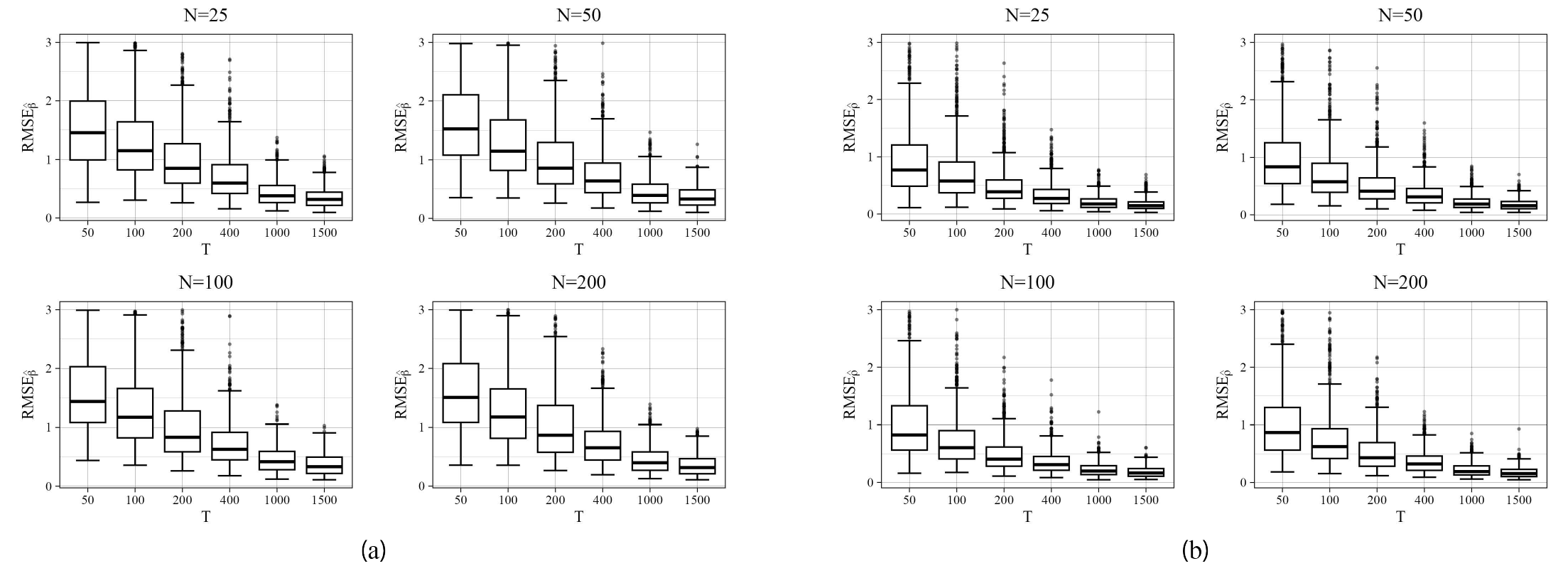}
  \caption{Boxplots of the joint convergence error for model (\ref{YW:lse}) with $\lambda = 0$. The statistics are defined in Figure~\ref{fig:thm2_1}, based on the first line of (\ref{rmsebt}).}\label{fig:thm2_2}.
\end{figure}

We first examine the joint convergence properties of \( \hat{\bbeta}_i \) established in \autoref{thm1}.  To evaluate its overall estimation accuracy, we use the root-mean-square error (RMSE), defined as
\begin{equation}\label{rmsebt}
    \mathrm{RMSE}_{\hat{\beta}} = \left\{
\begin{array}{ll}
\left( \frac{1}{N} \sum_{i=1}^N \| (\wh\bX_i' \wh\bX_i) (\wh\bbeta_i - \bbeta_i) \|_2^2 \right)^{1/2}, & \text{if } \lambda_i \to 0 \\
\left( \frac{1}{N} \sum_{i=1}^N \| \widehat{\bbeta}_i(\lambda_i) - \widehat{\bX}_i(\lambda_i)^{-1} \widehat{\bX}_i \widehat{\bX}_i' \bbeta_i \|_2^2 \right)^{1/2}, & \text{\textit{otherwise}.}
\end{array} \right.
\end{equation}
Here,  $\lambda_i$  is the ridge penalty parameter applied to the Yule-Walker equations for each sample.  We examine two cases: a relatively large  $\lambda_i = 10^{-3}$ and a much smaller $\lambda_i = 10^{-9}$. When  $\lambda_i \to 0$ (e.g., $\lambda_i = 10^{-9}$), the estimator closely resembles 
 that of the ordinary least squares (OLS) estimation, but we set $\lambda_i = 10^{-9}$ to avoid  singularity of \( (\wh\bX_i' \wh\bX_i) \).  
Figure~\ref{fig:thm2_1}(a) and (b) present the boxplots of the RMSEs of $\wh\bbeta(\lambda_i)$'s (denoted by RMSE$_{\wh\bbeta}$) and $\wh\rho_i(\lambda_i)$'s (denoted by RMSE$_{\wh\rho}$), respectively, computed using the second formula in (\ref{rmsebt}). From Figure~\ref{fig:thm2_1}, we see that the $\mathrm{RMSE}_{\hat{\beta}}$ and RMSE$_{\wh\rho}$ decrease as the sample size $T$ increases, which is in agreement with the  theoretical results in Theorem~\ref{thm1}. Similar patterns can also be found in Figure~\ref{fig:thm2_2} for $\lambda\rightarrow 0$ using the RMSE defined in the first line of (\ref{rmsebt}).


\begin{figure}[!ht]
  \includegraphics[width = \textwidth]{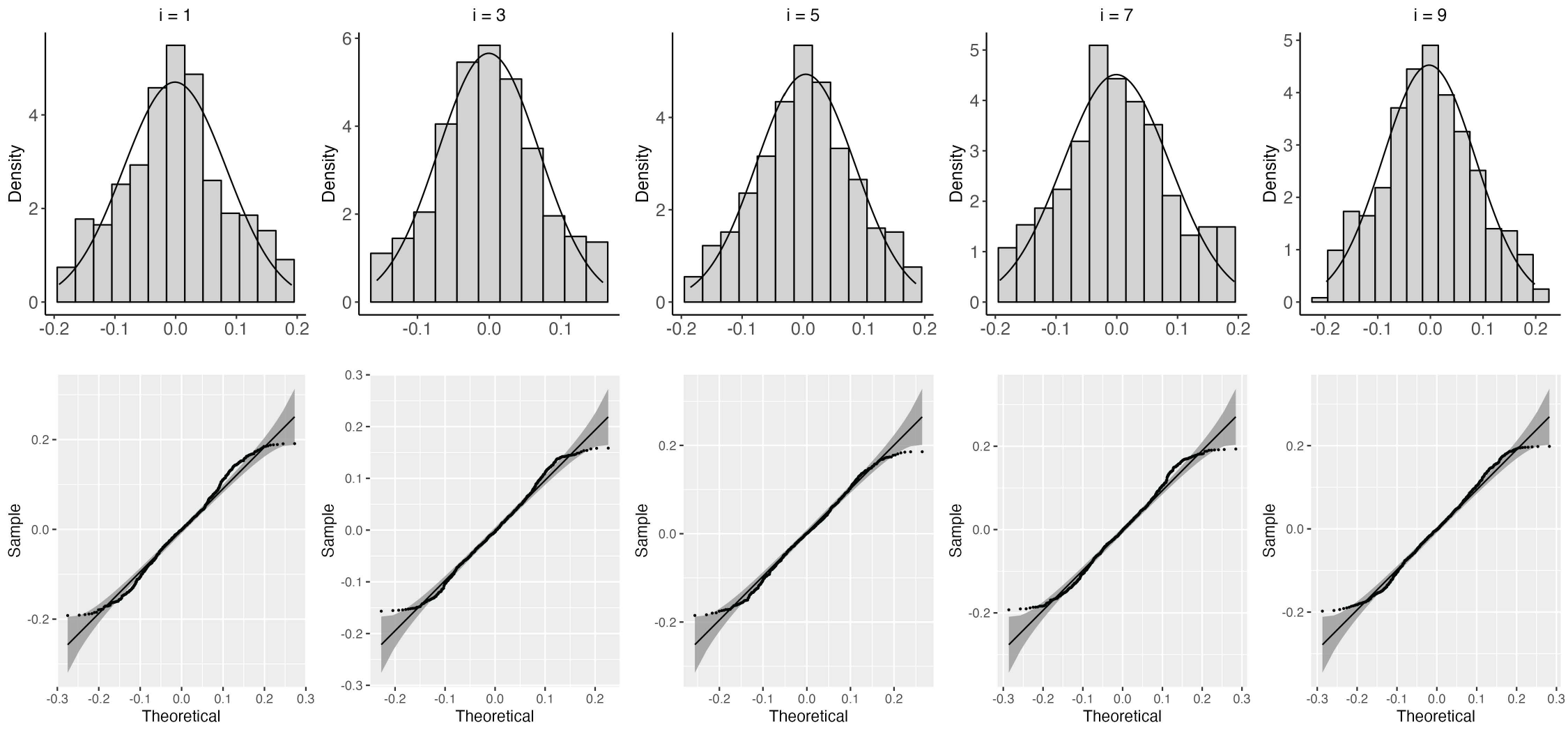}
  \caption{Histograms of the five spatial coefficient estimates and their corresponding empirical and theoretical distribution plots for Eq~(\ref{YW:lse}).The histograms show the distribution of the first component of $\bV_i(\wh\bbeta_i-\bbeta_i)$ over 1000 iterations, under the setting of $N = 25$ and $T = 1500$. The superimposed normal curve represents the theoretical distribution from Theorem~\ref{thm2}, with mean 0 and variance given by the first diagonal element of $\bX_i'\bU_i\bX_i$. Here, $i = 1, 3, 5, 7, 9$ correspond to the 1st, 3rd, 5th, 7th, and 9th component in a dataset of dimension $N=25$. 
  }\label{fig:thm3}
\end{figure}

\begin{table}[!htbp]
\centering
\caption{Coverage rates for $\bV_i(\wh\bbeta_i-\bbeta_i)$ across different significance levels. }
\label{tab:thm3}
\renewcommand\arraystretch{1.3}
\resizebox{\textwidth}{!}{%
\begin{tabular}{ccccccc}
\hline
\textbf{Significance} & \multicolumn{6}{c}{\textbf{Coverage}}                                                       \\ \hline
\textbf{0.1}          & 0.801(0.013) & 0.852(0.011) & 0.941(0.007)  & 0.966(0.006)  & 0.989(0.003)  & 0.998(0.001)  \\
\textbf{0.05}         & 0.842(0.012) & 0.891(0.010) & 0.957(0.006)  & 0.983(0.004)  & 0.997(0.002)  & 1.000(0.000)  \\
\textbf{0.01}         & 0.898(0.010) & 0.935(0.008) & 0.980(0.004)  & 0.993(0.003)  & 0.998(0.001)  & 1.000(0.000)  \\ \hline
\textbf{$T$}          & \textbf{250} & \textbf{500} & \textbf{1000} & \textbf{2000} & \textbf{3000} & \textbf{5000} \\ \hline
\end{tabular}%
}
\captionsetup{font=small}
\caption*{\textit{Note:} This table shows the coverage rate of the first component of $\bV_1(\wh\bbeta_1-\bbeta_1)$ within the confidence intervals from the theoretical distribution in Theorem \ref{thm2}. The theoretical distribution has a zero mean and a variance equal to the first diagonal element of $\bX_1'\bU_1\bX_1$. $T$ is the sample size. The results are based on 1000 iterations with a cross-sectional dimension $N = 25$.
}
\end{table}

To assess the distributional properties of the estimates in Theorem~\ref{thm2}, we present histograms of the first component of \(\wh\bV_i(\wh\bbeta_i - \bbeta_i)\), for $i=1,3,5,7$ and $9$ in \autoref{fig:thm3},  along with their theoretical density curves computed from the limiting distribution in Theorem~\ref{thm2}, where $\wh\bV_i$ is the sample estimator for $\bV_i$ defined in Eq~(\ref{Vi}). The histograms and their corresponding QQ-plots in Figure~\ref{fig:thm3} suggest that the entries of $\wh\bV_i(\wh\bbeta_i - \bbeta_i)$ asymptotically follow a normal distribution, which aligns with our theoretical results. Furthermore, in \autoref{tab:thm3}, we evaluate the asymptotic properties by reporting the coverage rates of the estimators in Eq~(\ref{YW:lse}) under varying significance levels and sample sizes 
$T$. As 
$T$ increases, the coverage rates exhibit a clear improvement, consistent with the theoretical results in \autoref{thm2}.

As a shrinkage-based approach, it is interesting to directly assess the estimation accuracy of \(\bbeta_i\). We define the coefficient error (CE) as
\begin{equation}
    \mathrm{CE}_{\hat{\beta}} = \left( \frac{1}{N} \sum_{i=1}^N \| (\wh\bbeta_i(\lambda_i) - \bbeta_i) \|_2^2 \right)^{1/2},
\end{equation}
which measures the deviation of \(\wh\bbeta_i\) from the true parameter \(\bbeta_i\), thereby capturing the overall estimation error.
\autoref{tab:known_k1k0} reports the coefficient error for the ridge regression estimates $\wh\bbeta_i(\lambda_i)$ when factors are observed. 
The results indicate that ridge regression ($\lambda_i = 10^{-3}$) yields lower error and variance compared to OLS ($\lambda_i = 10^{-9}$). Additionally, the estimation errors are reduced when stacking the cases of 
$k=0$ and 
$k=1$ together in the Yule-Walker equations, compared to relying solely on 
$k=0$.


\begin{table}[!ht]
\centering
\caption{Comparison of CE for ridge regression estimators with observed factors across different penalized parameter $\lambda_i$  and the lagging factor impact.}
\label{tab:known_k1k0}
\renewcommand\arraystretch{1.4}
\resizebox{\textwidth}{!}{%
\begin{tabular}{ccccccccccc}
\hline
                     &      &  & \multicolumn{2}{c}{$k =1\enspace (\lambda_i=10^{-9})$}   &  & \multicolumn{2}{c}{$k=1\enspace (\lambda_i=10^{-3})$}    &  & \multicolumn{2}{c}{$k=0\enspace (\lambda_i =10^{-3})$}   \\ \cline{4-5} \cline{7-8} \cline{10-11} 
N                    & T    &  & $\mathrm{CE}_{\hat{\beta}}$ & $\mathrm{CE}_{\hat{\rho}}$ &  & $\mathrm{CE}_{\hat{\beta}}$ & $\mathrm{CE}_{\hat{\rho}}$ &  & $\mathrm{CE}_{\hat{\beta}}$ & $\mathrm{CE}_{\hat{\rho}}$ \\ \hline
\multirow{6}{*}{25}  & 50   &  & 1.394(0.495)                & 0.291(0.276)               &  & 1.021(0.263)                & 0.135(0.102)               &  & 1.251(0.122)                & 0.165(0.132)               \\
                     & 100  &  & 1.405(0.510)                & 0.357(0.326)               &  & 1.012(0.260)                & 0.124(0.101)               &  & 1.238(0.129)                & 0.164(0.136)               \\
                     & 200  &  & 1.444(0.509)                & 0.488(0.379)               &  & 0.971(0.235)                & 0.120(0.097)               &  & 1.239(0.125)                & 0.167(0.131)               \\
                     & 400  &  & 1.562(0.533)                & 0.662(0.399)               &  & 0.936(0.220)                & 0.117(0.096)               &  & 1.233(0.124)                & 0.171(0.137)               \\
                     & 1000 &  & 1.810(0.522)                & 0.925(0.358)               &  & 0.976(0.219)                & 0.133(0.113)               &  & 1.236(0.122)                & 0.175(0.137)               \\
                     & 1500 &  & 1.906(0.522)                & 0.990(0.347)               &  & 1.000(0.210)                & 0.129(0.104)               &  & 1.231(0.125)                & 0.171(0.136)               \\ \hline
\multirow{6}{*}{50}  & 50   &  & 1.424(0.444)                & 0.328(0.248)               &  & 1.009(0.210)                & 0.154(0.089)               &  & 1.219(0.086)                & 0.203(0.124)               \\
                     & 100  &  & 1.429(0.445)                & 0.412(0.307)               &  & 0.992(0.206)                & 0.144(0.082)               &  & 1.218(0.085)                & 0.200(0.124)               \\
                     & 200  &  & 1.444(0.457)                & 0.525(0.333)               &  & 0.934(0.195)                & 0.131(0.073)               &  & 1.213(0.089)                & 0.202(0.124)               \\
                     & 400  &  & 1.587(0.479)                & 0.713(0.349)               &  & 0.900(0.182)                & 0.130(0.079)               &  & 1.215(0.084)                & 0.202(0.124)               \\
                     & 1000 &  & 1.838(0.460)                & 0.947(0.310)               &  & 0.931(0.174)                & 0.133(0.079)               &  & 1.205(0.087)                & 0.192(0.116)               \\
                     & 1500 &  & 1.918(0.445)                & 1.010(0.290)               &  & 0.966(0.170)                & 0.144(0.089)               &  & 1.208(0.086)                & 0.198(0.125)               \\ \hline
\multirow{6}{*}{100} & 50   &  & 1.467(0.415)                & 0.318(0.239)               &  & 1.010(0.181)                & 0.125(0.053)               &  & 1.177(0.057)                & 0.172(0.077)               \\
                     & 100  &  & 1.471(0.400)                & 0.407(0.271)               &  & 0.995(0.173)                & 0.120(0.048)               &  & 1.172(0.059)                & 0.168(0.075)               \\
                     & 200  &  & 1.502(0.414)                & 0.556(0.325)               &  & 0.916(0.158)                & 0.104(0.043)               &  & 1.175(0.061)                & 0.163(0.074)               \\
                     & 400  &  & 1.591(0.445)                & 0.725(0.324)               &  & 0.847(0.152)                & 0.101(0.041)               &  & 1.172(0.060)                & 0.169(0.075)               \\
                     & 1000 &  & 1.829(0.425)                & 0.937(0.284)               &  & 0.877(0.148)                & 0.106(0.053)               &  & 1.174(0.059)                & 0.166(0.080)               \\
                     & 1500 &  & 1.920(0.386)                & 1.008(0.245)               &  & 0.913(0.143)                & 0.113(0.049)               &  & 1.171(0.060)                & 0.171(0.079)               \\ \hline
\multirow{6}{*}{200} & 50   &  & 1.443(0.431)                & 0.273(0.238)               &  & 0.988(0.195)                & 0.086(0.042)               &  & 1.174(0.038)                & 0.120(0.064)               \\
                     & 100  &  & 1.434(0.430)                & 0.352(0.283)               &  & 0.969(0.184)                & 0.080(0.040)               &  & 1.169(0.041)                & 0.114(0.062)               \\
                     & 200  &  & 1.458(0.436)                & 0.505(0.322)               &  & 0.897(0.171)                & 0.072(0.036)               &  & 1.167(0.041)                & 0.114(0.061)               \\
                     & 400  &  & 1.566(0.462)                & 0.680(0.346)               &  & 0.843(0.150)                & 0.068(0.034)               &  & 1.165(0.041)                & 0.113(0.061)               \\
                     & 1000 &  & 1.834(0.453)                & 0.938(0.303)               &  & 0.868(0.145)                & 0.072(0.037)               &  & 1.165(0.042)                & 0.116(0.062)               \\
                     & 1500 &  & 1.910(0.427)                & 0.995(0.274)               &  & 0.909(0.140)                & 0.076(0.039)               &  & 1.167(0.041)                & 0.117(0.062)               \\ \hline
\end{tabular}%
}
\caption*{\textit{Note:}  Here, $k=1$  represents the application of the stacking strategy, which incorporates lagged factors $\bff_{t-1}$ as instrumental variables, whereas  $k=0$  indicates the stacking strategy is not applied, relying exclusively on the contemporaneous factors $\bff_t$, as detailed in Section~\ref{sec22}.}
\end{table}

\begin{figure}[!ht]
  \includegraphics[width = \textwidth]{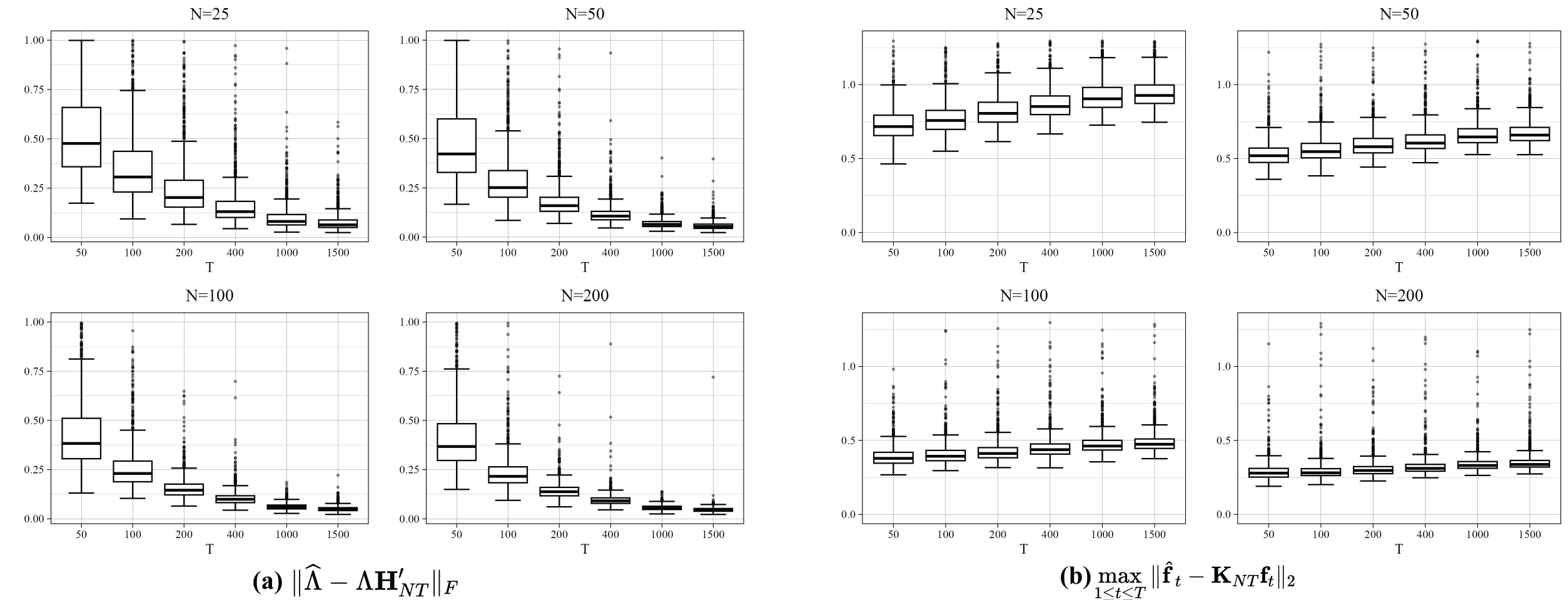}
  \caption{Boxplots of the convergence performance for the estimated loading matrices and latent factors in Eq~(\ref{ft:latent}).}\label{fig:thm45}
\end{figure}

Next, we investigate the performance of our proposed method in scenarios where the common factors are unobserved, focusing on the recovery of latent factors $\wh\bff_t$ and the estimation accuracy of the parameters $\brho$ and $\bB$ in Eq~(\ref{ft:latent}). \autoref{fig:thm45} illustrates the convergence behavior of the estimated loading matrix and latent factors using our method. In Panel (a) of \autoref{fig:thm45}, it is evident that the loading matrix converges steadily as $T$ increases. On the other hand, for each fixed $N$, Panel (b) reveals that $\max_{1 \leq t \leq T} \|\wh\bff_t - \bK_{NT} \bff_t\|_2$ increases with $T$, which is reasonable since the uniform distance is measured over the entire $T$-period. However, for each fixed $T$, the uniform distance will become smaller as $N$ increases, which is in agreement with our theoretical results. 

\begin{figure}[!ht]
  \includegraphics[width = \textwidth]{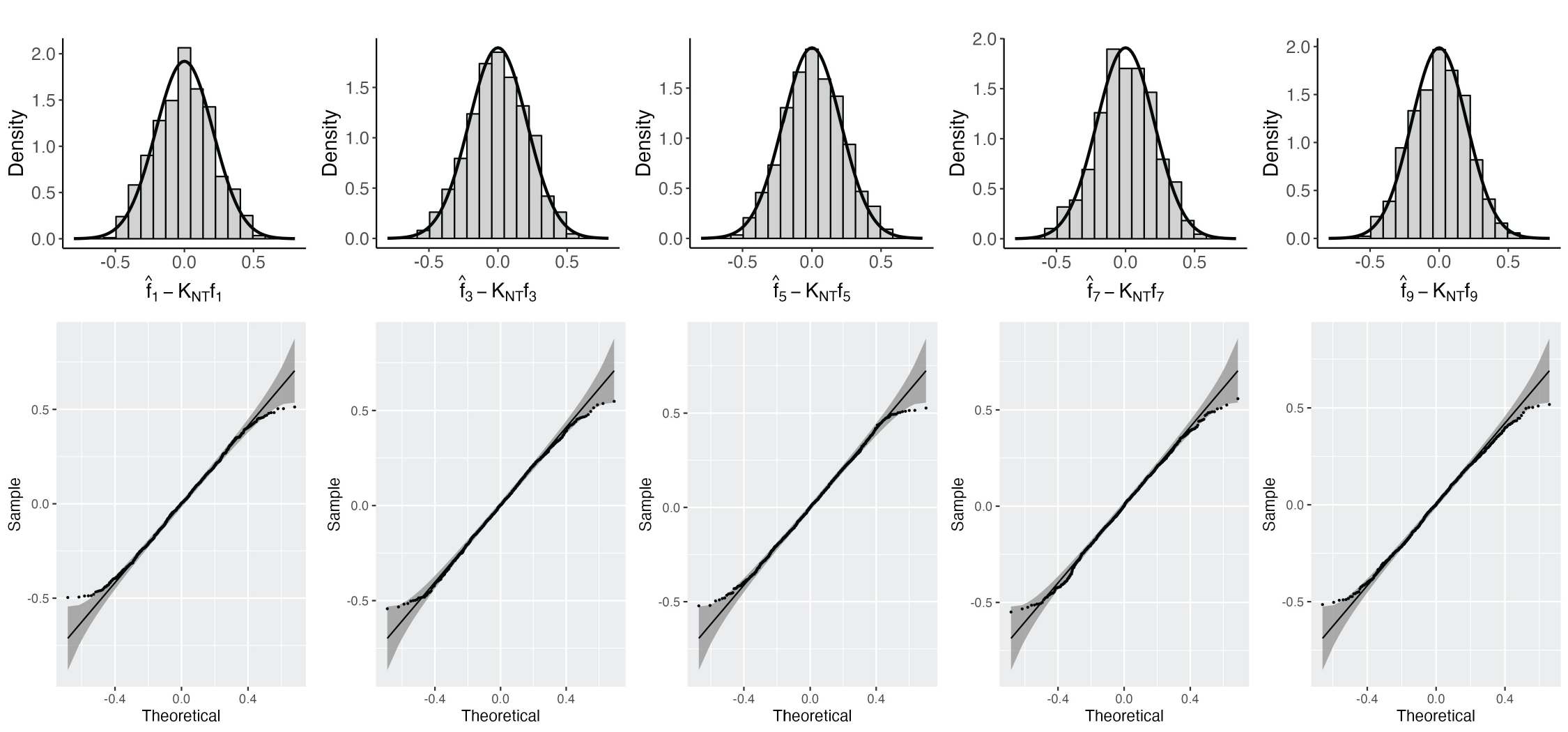}
  \caption{Histograms of $\wh\bff_t - \bK_{NT}\bff_t$ and their corresponding
empirical and theoretical distribution plots for Model~(\ref{ft:latent}). The results are based on 1000 iterations, focusing on the first element of \(\wh\bff_t - \bK_{NT}\bff_t\), for t = 1, 3, 5, 7, 9. The superimposed normal curves represent the theoretical distribution derived in Theorem~\ref{thm4}(ii), with mean 0 and variance given by the first diagonal element of \(\bH\bGamma_t\bH'\). The simulation results are obtained under $(N, T) = (25, 1500)$.}\label{fig:thm5}
\end{figure}

Now, we examine the asymptotic normality of the estimated latent factors. \autoref{fig:thm5} presents the histograms and QQ-plots of the first element of \(\wh\bff_t - \bK_{NT} \bff_t\) for $t=1,3,5,7$ and $9$ when $T=1500$, which clearly show an asymptotic normality pattern across all settings. In addition, \autoref{tab:thm5} presents the average coverage rates of the first component of  \(\wh\bff_t - \bK_{NT} \bff_t\) for $t\in\{1,6,11,...,96\}$, with a total of 20 factors, across different significance levels and values of $N$. As $N$ increases, the average coverage rates gradually improve, accompanied by reduced variance. These findings align with \autoref{thm3} and \autoref{thm4}.

\begin{table}[!ht]
\centering
\caption{Coverage rates of  $\wh\bff_t - \bK_{NT} \bff_t$  across different significance levels}
\label{tab:thm5}
\renewcommand\arraystretch{1.3}
\resizebox{0.85\textwidth}{!}{%
\begin{tabular}{cllll}
\hline
\textbf{Significance} & \multicolumn{4}{c}{\textbf{Coverage}}                                                                                             \\ \hline
\textbf{0.1}                & 0.910(0.0090)                    & 0.922(0.0085)                    & 0.943(0.0073)                     & 0.987(0.0036)                      \\
\textbf{0.05}               & 0.957(0.0064)                    & 0.959(0.0063)                    & 0.974(0.0059)                     & 0.993(0.0026)                      \\
\textbf{0.01}               & 0.991(0.0030)                    & 0.991(0.0030)                    & 0.994(0.0024)                     & 0.998(0.0014)                      \\ \hline
\textbf{$N$}                  & \multicolumn{1}{c}{\textbf{25}} & \multicolumn{1}{c}{\textbf{50}} & \multicolumn{1}{c}{\textbf{100}} & \multicolumn{1}{c}{\textbf{200}} \\ \hline
\end{tabular}%
}
\caption*{\textit{Note:} This table shows the coverage rate of the first component of $\wh\bff_t-\bK_{NT}\bff_t$ within the confidence intervals from the theoretical distribution in Theorem~\ref{thm4}(ii). The theoretical distribution has mean zero and variance given by the first diagonal element of $\bH\bGamma_t\bH'$. The results are based on 1000 iterations with sample size $T=1500$.}
\end{table}

\begin{figure}[!ht]
  \includegraphics[width = \textwidth]{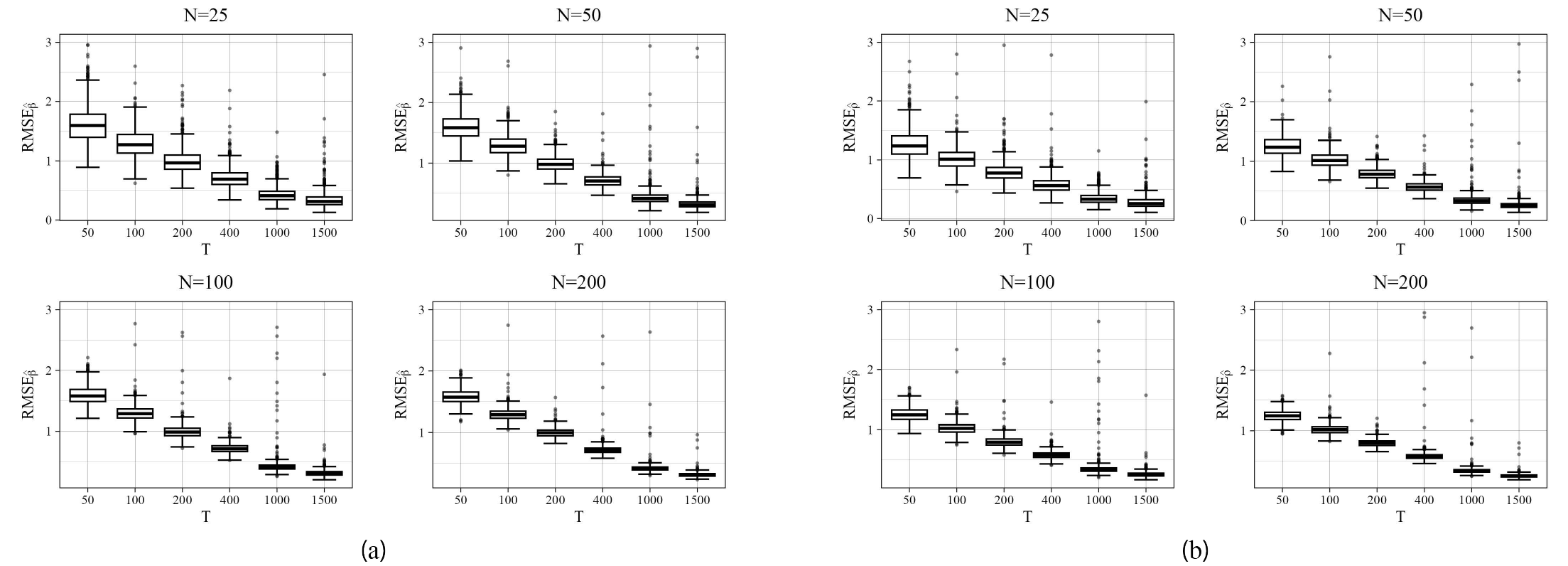}
  \caption{Boxplots of estimator convergence for model (\ref{yw:latent}) with fixed small $\lambda_i$. }\label{fig:thm6_1}
\end{figure}
\begin{figure}[!ht]
  \includegraphics[width = \textwidth]{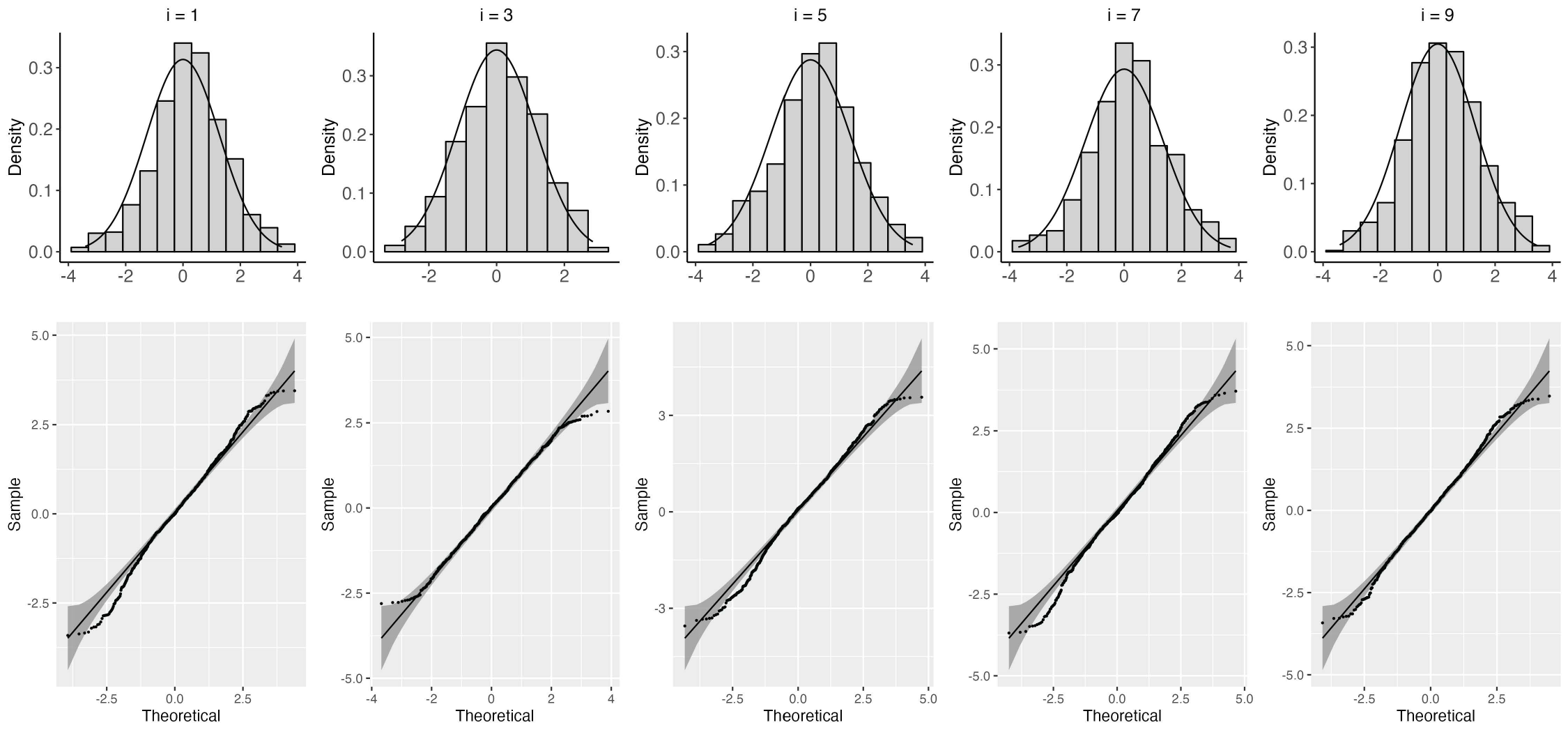}
  \caption{Histograms of the five spatial coefficient estimates and their corresponding empirical and theoretical distribution plots for model~(\ref{yw:latent}). The histograms show the distribution of the first component of $\bV_i^H(\wt\bbeta_i-\bK_{NT}^*\bbeta_i)$ as defined in \autoref{thm5}. The superimposed normal curve represents the theoretical distribution from Theorem~\ref{thm5}. Here, $i = 1, 3, 5, 7, 9$ correspond to the 1st, 3rd, 5th, 7th, and 9th samples in a dataset of size $N$. The results are based on 1,000 iterations with $(N, T) = (25, 3000)$.
  }\label{fig:thm6_2}
\end{figure}

With the estimated factors $\wh\bff_t$, 
\autoref{fig:thm6_1} demonstrates the boxplots of the RMSE of $\wt\bbeta_i$ under increasing $T$, where the RMSE is similarly define as (\ref{rmsebt}). The patterns of the RMSEs in \autoref{fig:thm6_1} are similar to those in Figure~\ref{fig:thm2_2}, and we omit the details here. To validate the distributional properties, \autoref{fig:thm6_2} displays the histogram and QQ-plot of the first component of $\bV_i^H(\wt\bbeta_i-\bK_{NT}^*\bbeta_i)$, for $i=1,3,5,7$ and $9$. From Figure~\ref{fig:thm6_1}, we can see clearly an asymptotic normality pattern across all settings, which is in line with our theoretical results in Theorem~\ref{thm5}. Moreover, we verify the coverage probabilities of the first component of $\bV_1^H(\wt\bbeta_1-\bK_{NT}^*\bbeta_1)$ in Table~\ref{fig:thm6_2}, which are also in agreement with our theory.

Next, we present the coefficient error (CE) results for ridge estimators in the case of unknown factors in \autoref{tab:unknown_k1k0}. From \autoref{tab:unknown_k1k0} we see that integrating the stacking strategy with proper ridge penalty improves the estimation accuracy, further validating the proposed method.

\begin{table}[!ht]
\centering
\caption{Coverage performance of $\bV_i^H(\wt\bbeta_i-\bK_{NT}^*\bbeta_i)$ under latent factor estimation across different significance levels.}
\label{tab:thm6}
\renewcommand\arraystretch{1.4}
\resizebox{\textwidth}{!}{%
\begin{tabular}{ccccccccc}
\hline
\textbf{Significance} & \multicolumn{8}{c}{\textbf{Coverage}}                                                                                     \\ \hline
\textbf{0.1}          & 0.343(0.016) & 0.438(0.017) & 0.531(0.017) & 0.663(0.016) & 0.819(0.013)  & 0.932(0.008)  & 0.979(0.005)  & 1.000(0.000)  \\
\textbf{0.05}         & 0.399(0.016) & 0.507(0.017) & 0.595(0.017) & 0.718(0.015) & 0.866(0.011)  & 0.974(0.005)  & 1.000(0.000)  & 1.000(0.000)  \\
\textbf{0.01}         & 0.477(0.017) & 0.595(0.017) & 0.692(0.016) & 0.786(0.014) & 0.946(0.008)  & 1.000(0.000)  & 1.000(0.000)  & 1.000(0.000)  \\ \hline
\textbf{$T$}          & \textbf{50}  & \textbf{100} & \textbf{200} & \textbf{400} & \textbf{1000} & \textbf{2000} & \textbf{3000} & \textbf{5000} \\ \hline
\end{tabular}%
}
\caption*{\textit{Note:} This table shows the coverage rate of the first component of $\bV_1^H(\wt\bbeta_1-\bK_{NT}^*\bbeta_1)$, illustrating the asymptotic performance of $\wt\rho_1$ within the confidence intervals derived from the theoretical distribution,  based on 1,000 iterations with $N = 25$. The theoretical distribution has zero mean and variance corresponding to the first diagonal element of $\bX_1^H{'}\bU_1^H\bX_1^H$, which is defined in \autoref{thm5}.}
\end{table}

\begin{table}[!ht]
\centering
\caption{Comparison of coefficient error (CE) for ridge regression estimators with latent factors across different penalized parameter $\lambda_i$ and the lagging factor impact.}
\label{tab:unknown_k1k0}
\renewcommand\arraystretch{1.4}
\resizebox{\textwidth}{!}{%
\begin{tabular}{ccccccccccc}
\hline
                     &      &  & \multicolumn{2}{c}{$k =1\enspace (\lambda_i=10^{-9})$}   &  & \multicolumn{2}{c}{$k=1\enspace (\lambda_i=1)$}          &  & \multicolumn{2}{c}{$k=0\enspace (\lambda_i =1)$}         \\ \cline{4-5} \cline{7-8} \cline{10-11} 
N                    & T    &  & $\mathrm{CE}_{\hat{\beta}}$ & $\mathrm{CE}_{\hat{\rho}}$ &  & $\mathrm{CE}_{\hat{\beta}}$ & $\mathrm{CE}_{\hat{\rho}}$ &  & $\mathrm{CE}_{\hat{\beta}}$ & $\mathrm{CE}_{\hat{\rho}}$ \\ \hline
\multirow{6}{*}{25}  & 50   &  & 2.417(0.139)                & 1.201(0.184)               &  & 2.336(0.122)                & 1.072(0.235)               &  & 2.339(0.142)                & 1.080(0.251)               \\
                     & 100  &  & 2.410(0.152)                & 1.194(0.202)               &  & 2.322(0.130)                & 1.068(0.247)               &  & 2.330(0.143)                & 1.066(0.263)               \\
                     & 200  &  & 2.394(0.166)                & 1.190(0.178)               &  & 2.324(0.128)                & 1.076(0.236)               &  & 2.327(0.141)                & 1.077(0.242)               \\
                     & 400  &  & 2.396(0.178)                & 1.173(0.183)               &  & 2.314(0.128)                & 1.073(0.232)               &  & 2.336(0.131)                & 1.090(0.241)               \\
                     & 1000 &  & 2.419(0.144)                & 1.199(0.173)               &  & 2.310(0.143)                & 1.070(0.244)               &  & 2.334(0.139)                & 1.080(0.260)               \\
                     & 1500 &  & 2.435(0.153)                & 1.195(0.190)               &  & 2.323(0.126)                & 1.078(0.234)               &  & 2.323(0.139)                & 1.080(0.247)               \\ \hline
\multirow{6}{*}{50}  & 50   &  & 2.346(0.142)                & 1.001(0.230)               &  & 2.262(0.150)                & 0.838(0.306)               &  & 2.288(0.117)                & 0.973(0.277)               \\
                     & 100  &  & 2.334(0.119)                & 1.062(0.179)               &  & 2.267(0.134)                & 0.937(0.270)               &  & 2.302(0.104)                & 1.041(0.232)               \\
                     & 200  &  & 2.325(0.100)                & 1.055(0.175)               &  & 2.259(0.134)                & 0.905(0.281)               &  & 2.282(0.105)                & 0.998(0.268)               \\
                     & 400  &  & 2.362(0.123)                & 1.087(0.162)               &  & 2.288(0.134)                & 0.937(0.263)               &  & 2.306(0.114)                & 1.013(0.254)               \\
                     & 1000 &  & 2.359(0.106)                & 1.084(0.167)               &  & 2.285(0.122)                & 0.920(0.254)               &  & 2.300(0.104)                & 1.027(0.239)               \\
                     & 1500 &  & 2.362(0.108)                & 1.087(0.163)               &  & 2.278(0.136)                & 0.926(0.262)               &  & 2.288(0.115)                & 1.029(0.225)               \\ \hline
\multirow{6}{*}{100} & 50   &  & 2.244(0.122)                & 0.828(0.243)               &  & 2.219(0.135)                & 0.797(0.291)               &  & 2.289(0.108)                & 1.057(0.258)               \\
                     & 100  &  & 2.261(0.102)                & 0.912(0.232)               &  & 2.227(0.119)                & 0.767(0.302)               &  & 2.299(0.097)                & 1.033(0.266)               \\
                     & 200  &  & 2.248(0.126)                & 0.931(0.237)               &  & 2.220(0.127)                & 0.835(0.289)               &  & 2.317(0.075)                & 1.115(0.179)               \\
                     & 400  &  & 2.265(0.113)                & 0.947(0.223)               &  & 2.245(0.128)                & 0.880(0.304)               &  & 2.309(0.094)                & 1.081(0.228)               \\
                     & 1000 &  & 2.278(0.091)                & 0.965(0.194)               &  & 2.242(0.119)                & 0.929(0.266)               &  & 2.314(0.071)                & 1.139(0.185)               \\
                     & 1500 &  & 2.267(0.102)                & 0.939(0.214)               &  & 2.246(0.124)                & 0.870(0.291)               &  & 2.308(0.092)                & 1.089(0.208)               \\ \hline
\multirow{6}{*}{200} & 50   &  & 2.230(0.116)                & 0.864(0.282)               &  & 2.230(0.086)                & 0.886(0.230)               &  & 2.352(0.055)                & 1.213(0.148)               \\
                     & 100  &  & 2.228(0.098)                & 0.890(0.232)               &  & 2.217(0.101)                & 0.849(0.262)               &  & 2.345(0.066)                & 1.218(0.166)               \\
                     & 200  &  & 2.247(0.096)                & 0.945(0.248)               &  & 2.230(0.098)                & 0.921(0.231)               &  & 2.350(0.047)                & 1.230(0.104)               \\
                     & 400  &  & 2.251(0.096)                & 0.950(0.233)               &  & 2.231(0.110)                & 0.918(0.250)               &  & 2.333(0.084)                & 1.184(0.231)               \\
                     & 1000 &  & 2.251(0.086)                & 0.970(0.214)               &  & 2.217(0.104)                & 0.880(0.271)               &  & 2.331(0.080)                & 1.180(0.242)               \\
                     & 1500 &  & 2.254(0.094)                & 0.978(0.223)               &  & 2.259(0.093)                & 0.986(0.237)               &  & 2.339(0.093)                & 1.198(0.228)               \\ \hline
\end{tabular}%
}
\end{table}

Finally, we compare the predictive performance of our method with QMLE by evaluating their out-of-sample forecasting accuracy under heterogeneous conditions. The forecasting error (FE) is defined as
 \begin{equation}
    \mathrm{FE} = \left( \frac{1}{N(T-T_1)} \sum_{t=T_1+1}^{T} \| (\wh \by_t - \by_t) \|_2^2 \right)^{1/2},
\end{equation}
where $\wh \by_t$ denotes the predicted value using the estimated coefficients from the training sample, and $\by_t$ represents the actual value. \autoref{tab:sim-pre} presents the forecasting error and standard deviation of both method across different cross-sectional dimensions ($N$), with the out-of-sample period set from $T_1+1=321$ and $T_1+1=400$. 
The proposed model with lagged factor instruments $(k=1)$ achieves lower forecast error across all 
$N$ dimensions, outperforming QMLE. This results aligns with prior simulations that emphasize the benefits of combining shrinkage techniques with lagging factor integration to enhance accuracy.

\begin{table}[!ht]
\centering
\caption{Out-of-sample simulation evaluation of different models with best ones in boldface.}
\label{tab:sim-pre}
\renewcommand\arraystretch{1.2}
\resizebox{\textwidth}{!}{%
\begin{tabular}{cccccccc}
\hline
    & \multicolumn{2}{c}{Proposed Model $(\lambda_i=10^{-3})$} &  & \multicolumn{2}{c}{Proposed Model $(\lambda_i=10^{-9})$} &  & \multirow{2}{*}{QML} \\ \cline{2-3} \cline{5-6}
$N$ & $k=0$                  & $k=1$                           &  & $k=0$                       & $k=1$                      &  &                      \\ \hline
25  & 2.672(0.714)           & \textbf{1.046(0.087)}           &  & 2.605(0.689)                & 1.047(0.088)               &  & 1.900(0.399)         \\
50  & 2.750(1.043)           & \textbf{1.047(0.090)}           &  & 2.618(0.978)                & 1.048(0.091)               &  & 1.577(0.353)         \\
100 & 2.925(1.589)           & \textbf{1.051(0.093)}           &  & 2.634(1.386)                & 1.055(0.096)               &  & 1.357(0.224)         \\
200 & 3.423(2.753)           & \textbf{1.055(0.101)}           &  & 2.666(1.996)                & 1.062(0.111)               &  & 1.353(0.347)         \\ \hline
\end{tabular}%
}
\captionsetup{font=small}
\caption*{\textit{Note:}  This table compares out-of-sample forecast errors between the proposed model and QML method under varying configurations of regularization parameters ($\lambda_i = 10^{-3}$ and $\lambda_i = 10^{-9}$) and lagging factor instruments ($k=0$ and $k=1$). The settings for the simulation include $T=400$ (time periods) and $K=3$ (factor dimensions).}
\end{table}

 \section{Empirical Studies}\label{sec5}
 
In this section, we apply the proposed method to two arbitrage pricing case studies. The first one focuses on modeling and forecasting stock returns of the S\&P 500 constituents, while the second examines quarterly changes in real housing prices across U.S. Metropolitan Statistical Areas (MSAs). For each case, we use the \texttt{R} software with a fixed random seed (\texttt{1234}) to randomly select a subset of cross-sectional units (denoted by $N$) and a subsample from the beginning of the full time span (denoted by $T$). The first $80\%$ of each subsample is used as the training set, and the remaining $20\%$ as the testing set for evaluating out-of-sample forecasting performance. We compare the proposed approach with the QMLE method and the classical Fama-French factor model without spatial interaction.



\subsection{Empirical Application to Stock Returns} \label{sec51}

Companies located in close geographic proximity often share exposure to regional policies and industrial clusters, highlighting the importance of accounting for spatial dependencies. In this example, we investigate how firm locations influence stock returns. Our analysis focuses on the daily log excess returns of S\&P 500 constituents from January 2004 to December 2016, comprising 3,273 time points across 205 companies. The companies are selected in the order provided by the original dataset, which is publicly available at \url{https://mpelger.people.stanford.edu/data-and-code}, with further details documented in \cite{pelger2020understanding}. As common factors, we incorporate the Fama-French variables, including the market factor (MKT), size factor (SMB), and value factor (HML).
To capture spatial dependence in each selected subsample with cross-sectional dimension $N$ and time dimension $T$, we construct a spatial weight matrix following a standard geographic approach from spatial econometrics. Specifically, we define an $N \times N$ spatial weight matrix $W$, where each off-diagonal entry $w_{ij}$ is given by $w_{ij} = (s_i d_{ij})^{-1}$ for $i \neq j$, and $w_{ii} = 0$. Here, $d_{ij}$ denotes the Haversine distance between the headquarters of companies $i$ and $j$, and $s_i = \sum_{j=1}^{N} d_{ij}^{-1}$ serves as a normalization factor to ensure that each row of $W$ sums to one. This construction ensures that the weights represent the relative geographic influence of company $j$ on company $i$.

Table~\ref{tab:emp51} presents a detailed comparison of the forecasting performance of our proposed method against QMLE and the classical Fama-French factor model across various configurations and industry classifications. As shown in the table, our method consistently achieves lower forecasting errors than both QMLE and the factor model, highlighting its effectiveness and the value of incorporating spatial dependencies into the arbitrage pricing framework.
In terms of computational efficiency, our method offers a substantial advantage over QMLE. For example, when $N = 200$ and $T = 1000$, QMLE requires over six hours on a standard CPU, while our approach produces comparable results in just a few minutes. This notable efficiency gain makes our method particularly attractive for large-scale applications, offering a favorable trade-off between accuracy and computational cost.


\begin{table}[!ht]
\centering
\caption{Forecast error comparison for stock returns with observed factors.}
\label{tab:emp51}
\renewcommand\arraystretch{1.8}
\resizebox{\textwidth}{!}{%
\begin{tabular}{ccccccccccccc}
\hline
\multicolumn{2}{c}{\multirow{2}{*}{\textbf{Method}}} & \multicolumn{3}{c}{Proposed method}                                                                                   &           & \multicolumn{3}{c}{QMLE}                                                                                              &  & \multicolumn{3}{c}{Factor model}                                                                                      \\ \cline{3-5} \cline{7-9} \cline{11-13} 
\multicolumn{2}{c}{}                                 & $N = 100$                                   & $N=150$                            & $N = 200$                          &           & $N = 100$                                   & $N = 150$                          & $N = 200$                          &  & $N = 100$                                   & $N=150$                            & $N = 200$                          \\ \hline
$T = 500$                                   &        & \textbf{0.8601 (0.2173)}                    & \textbf{0.8546 (0.2396)}           & \textbf{0.8462 (0.2510)}           & \textbf{} & 0.8665 (0.2177)                             & 0.8768 (0.2461)                    & 0.8772 (0.2653)                    &  & 0.8606 (0.2173)                             & 0.8551 (0.2396)                    & 0.8476 (0.2518)                    \\
$T = 1000$                                  &        & \textbf{0.9640 (0.1999)}                    & \textbf{0.9668 (0.2352)}           & \textbf{0.9866 (0.2791)}           & \textbf{} & 0.9861 (0.2038)                             & 1.0161 (0.2505)                    & 1.0542 (0.3073)                    &  & 0.9646 (0.2005)                             & 0.9678 (0.2360)                    & 0.9873 (0.2801)                    \\
$T = 2000 $                                 &        & \textbf{0.5630 (0.0641)}                    & \textbf{0.5704 (0.0745)}           & \textbf{0.5623 (0.0818)}           &           & 0.6042 (0.0844)                             & 0.6439 (0.1156)                    & 0.6723 (0.1510)                    &  & 0.5705 (0.0725)                             & 0.5821 (0.0863)                    & 0.5774 (0.0957)                    \\ \hline
\multirow{2}{*}{\textbf{GICS Class}}        &        & \multicolumn{3}{c}{Information Technology $(N=36)$}                                                                   &           & \multicolumn{3}{c}{Financials $(N=31)$}                                                                               &  & \multicolumn{3}{c}{Consumer Staples $(N=19)$}                                                                         \\ \cline{3-5} \cline{7-9} \cline{11-13} 
                                            &        & Proposed method                             & QMLE                               & Factor model                       &           & Proposed method                             & QMLE                               & Factor model                       &  & Proposed method                             & QMLE                               & Factor model                       \\ \hline
$T = 500 $                                  &        & \multicolumn{1}{l}{\textbf{0.7683(0.1371)}} & \multicolumn{1}{l}{0.8061(0.1431)} & \multicolumn{1}{l}{0.7701(0.1369)} & \textbf{} & \multicolumn{1}{l}{\textbf{0.7661(0.1223)}} & \multicolumn{1}{l}{0.7730(0.1256)} & \multicolumn{1}{l}{0.7669(0.1223)} &  & \multicolumn{1}{l}{\textbf{0.8765(0.1396)}} & \multicolumn{1}{l}{0.8766(0.1399)} & \multicolumn{1}{l}{0.8767(0.1398)} \\
$T = 1000$                                  &        & \multicolumn{1}{l}{\textbf{0.8338(0.1252)}} & \multicolumn{1}{l}{0.8697(0.1281)} & \multicolumn{1}{l}{0.8352(0.1249)} & \textbf{} & \multicolumn{1}{l}{\textbf{1.1530(0.2088)}} & \multicolumn{1}{l}{1.1535(0.2069)} & \multicolumn{1}{l}{1.1628(0.2123)} &  & \multicolumn{1}{l}{\textbf{0.9823(0.1024)}} & \multicolumn{1}{l}{0.9844(0.1021)} & \multicolumn{1}{l}{0.9837(0.1021)} \\
$T = 2000$                                  &        & \multicolumn{1}{l}{\textbf{0.6440(0.0726)}} & \multicolumn{1}{l}{0.6951(0.0804)} & \multicolumn{1}{l}{0.6441(0.0726)} &           & \multicolumn{1}{l}{\textbf{0.3859(0.0351)}} & \multicolumn{1}{l}{0.3963(0.0366)} & \multicolumn{1}{l}{0.3884(0.0351)} &  & \multicolumn{1}{l}{\textbf{0.6756(0.0556)}} & \multicolumn{1}{l}{0.6821(0.0565)} & \multicolumn{1}{l}{0.6793(0.0561)} \\ \hline
\end{tabular}%
}
\captionsetup{font=small}
\caption*{\textit{Note:} This table compares the forecast errors of the proposed model, QMLE, and the factor model across different combinations of $N$ and $T$, as well as three industry classifications: Information Technology, Financials, and Consumer Staples. These classifications are based on the Global Industry Classification Standard (GICS).}

\end{table}

Furthermore, we evaluate the forecasting errors for stock returns driven by unobserved factors, with the results summarized in Table~\ref{tab:emp51-2}. As shown in the table, the findings are consistent with our earlier results, further demonstrating the scalability and practicality of the proposed method for large-scale applications.

\begin{table}[!ht]
\centering
\caption{Forecast error comparison for stock returns with latent factors}
\label{tab:emp51-2}
\renewcommand\arraystretch{1.6}
\resizebox{\textwidth}{!}{%
\begin{tabular}{ccccccccccccc}
\hline
\multicolumn{2}{c}{\multirow{2}{*}{\textbf{Method}}} & \multicolumn{3}{c}{Proposed method}                                         &           & \multicolumn{3}{c}{QMLE}                            &  & \multicolumn{3}{c}{Factor model}                 \\ \cline{3-5} \cline{7-9} \cline{11-13} 
\multicolumn{2}{c}{}                                 & $N = 100$               & $N=150$                 & $N = 200$               &           & $N = 100$       & $N = 150$       & $N = 200$       &  & $N = 100$      & $N=150$        & $N = 200$      \\ \hline
$T = 500$                      &                     & \textbf{0.8218(0.2138)} & \textbf{0.8251(0.2353)} & \textbf{0.8157(0.2449)} & \textbf{} & 0.8468 (0.2132) & 0.8422 (0.2361) & 0.8353 (0.2491) &  & 0.8468(0.2137) & 0.8267(0.2346) & 0.8164(0.2444) \\
$T = 1000$                     &                     & \textbf{0.9092(0.1919)} & \textbf{0.9315(0.2261)} & \textbf{0.9250(0.2601)} & \textbf{} & 0.9369 (0.1968) & 0.9472 (0.2320) & 0.9605 (0.2669) &  & 0.9244(0.1931) & 0.9322(0.2257) & 0.9429(0.2591) \\
$T = 2000 $                    &                     & \textbf{0.5441(0.0690)} & \textbf{0.5501(0.0810)} & \textbf{0.5491(0.0902)} &           & 0.5560 (0.0704) & 0.5712 (0.0839) & 0.5698 (0.0938) &  & 0.5460(0.0695) & 0.5559(0.0823) & 0.5532(0.0913) \\ \hline
\end{tabular}%
}
\captionsetup{font=small}
\caption*{\textit{Note:} This table compares the forecast errors of the proposed model with QMLE and factor model under different combinations of $N$ and $T$. The number of latent factors is determined using the information criterion proposed by \cite{bai2002determining}.}
\end{table}
		
\subsection{Empirical Application to U.S. Housing Market}\label{sec52}

Our second application examines quarterly changes in real housing prices across 377 U.S. Metropolitan Statistical Areas (MSAs) from 1975-Q1 to 2014-Q4, as studied in \cite{aquaro2021estimation}. Due to shared supply and demand dynamics among neighboring regions, spatial models are essential for capturing such dependencies and enhancing predictive accuracy.

To account for broader economic influences—particularly the impact of stock market movements on real estate investment sentiment and capital allocation—we incorporate factor proxies from the previous example. For spatial dependence, we adopt the spatial weight matrix $W_{75}$ proposed in \cite{aquaro2021estimation}, in which MSAs within a specified radius $d$ are treated as neighbors (assigned a weight of 1), while non-neighbors receive a weight of 0. The resulting matrix is row-normalized to obtain the final weight matrix $W$.

Table~\ref{tab:emp52} presents a detailed comparison of forecasting errors for our proposed method, the QMLE approach from \cite{aquaro2021estimation}, and the Fama-French factor model without spatial interactions. As shown in the table, our method consistently yields lower forecast errors in most cases, demonstrating its robustness and efficiency across different data settings and reinforcing its applicability to spatial econometric forecasting.
	
\begin{table}[!ht]
\centering
\caption{Forecast error comparison for U.S. housing prices with observed factors.}
\label{tab:emp52}
\renewcommand\arraystretch{1.6}
\resizebox{\textwidth}{!}{%
\begin{tabular}{ccccccccccccc}
\hline
\multicolumn{2}{c}{\multirow{2}{*}{\textbf{Method}}} & \multicolumn{3}{c}{Proposed method}                                                 &           & \multicolumn{3}{c}{QMLE}                                                              &  & \multicolumn{3}{c}{Factor model}                                                    \\ \cline{3-5} \cline{7-9} \cline{11-13} 
\multicolumn{2}{c}{}                                 & \multicolumn{1}{c}{N = 20} & \multicolumn{1}{c}{N=50} & \multicolumn{1}{c}{N = 200} &           & \multicolumn{1}{c}{N = 20} & \multicolumn{1}{c}{N = 50} & \multicolumn{1}{c}{N = 200} &  & \multicolumn{1}{c}{N = 20} & \multicolumn{1}{c}{N=50} & \multicolumn{1}{c}{N = 200} \\ \hline
T = 50                       &                       & \textbf{1.2520(0.3110)}    & \textbf{1.6515(0.8584)}  & \textbf{1.5533(2.3174)}     & \textbf{} & 1.2680(0.2254)             & 1.6605(0.7631)             & 1.5863(1.5341)              &  & 1.2881(0.2231)             & 1.6931(0.7788)           & 1.7533(1.5479)              \\
T = 100                      &                       & \textbf{2.2001(0.6841)}    & \textbf{2.1557(1.0271)}  & \textbf{2.1723(2.4108)}     & \textbf{} & 2.2015(0.5007)             & 2.1577(0.7837)             & 2.1737(2.3700)              &  & 2.2163(0.4752)             & 2.1715(0.7257)           & 2.2269(1.4651)              \\
T = 150                      &                       & \textbf{2.8807(0.3411)}    & \textbf{2.8876(0.4750)}  & \textbf{2.7555(1.5827)}     &           & 2.8854(0.3358)             & 2.8910(0.4665)             & 2.7887(1.0758)              &  & 2.8870(0.3356)             & 2.8951(0.4645)           & 2.8076(1.0340)              \\ \hline
\end{tabular}%
}
\captionsetup{font=small}
\caption*{\textit{Note:} This table compares the forecast errors of the proposed model, QMLE and factor model in predicting U.S. housing prices, utilizing factors from the Fama-French three-factor model and a spatial weight matrix based on geometric distances.  }

\end{table}

In summary, the comparative analyses of S\&P 500 stock returns and U.S. housing prices demonstrate that our proposed method delivers superior predictive accuracy and computational efficiency, confirming its effectiveness across diverse spatial and temporal settings.

		\section{Conclusion} \label{sec6}
		
		This paper introduced a Spatial Arbitrage Pricing Theory (SAPT) model that integrates spatial interactions with multifactor structures involving both observable and latent variables. The SAPT framework offers two key conceptual innovations for asset pricing: (1) it introduces a spatial rho parameter, serving as a counterpart to the market beta in the classical CAPM; and (2) it captures spatial correlations typically unaccounted for in traditional Arbitrage Pricing Theory (APT) models, thereby extending the scope of standard CAPM and enhancing econometric tools for asset pricing analysis. For estimation, we proposed a generalized shrinkage Yule-Walker method that accommodates both observable and latent factors. The proposed methodology provides a flexible and computationally efficient framework for theoretical advancement and empirical research in financial and economic modeling.

        

		
	\end{onehalfspacing}

%
	
		\singlespacing
\bibliographystyle{econometrica-3}
\let\oldbibliography\thebibliography
\renewcommand{\thebibliography}[1]{%
  \oldbibliography{#1}%
  \setlength{\itemsep}{3pt}%
}
	{\footnotesize
		\bibliography{reference}
	}
	\onehalfspacing
%

\newpage

	
	\begin{titlepage}
	
	\begin{center}
	{\Large Online Appendix for\\High-Dimensional Spatial Arbitrage Pricing Theory with Heterogeneous Interactions}
	\end{center}

	 \thispagestyle{empty}
		\vspace{0.5cm}
		
		\begin{abstract}
			The online appendix collects the mathematical proofs  that support the main text. 
			\vspace{1cm}
			
			\noindent\textbf{Keywords:}  Spatial Arbitrage Pricing Theory, Multi-factor Analysis, Yule-Walker Estimation,  Eigenanalysis,  High Dimension

		\end{abstract}
	\end{titlepage}

	\setcounter{page}{1}
	
	\setcounter{section}{0}
	\setcounter{subsection}{0}
	
	\renewcommand{\thesection}{IA.\Alph{section}}
	\renewcommand{\thesubsection}{\thesection.\arabic{subsection}}
	
	\setcounter{equation}{0}
	\renewcommand{\theequation}{\thesection.\arabic{equation}}
	

	
	\renewcommand{\theequation}{IA.\arabic{equation}}%
	\renewcommand{\thefigure}{IA.\arabic{figure}} \setcounter{figure}{0}
	\renewcommand{\thetable}{IA.\Roman{table}} \setcounter{table}{0}

	\onehalfspacing      

	
	

\section{Proofs of the Theorems}

 We will use $C$ or $c$ to denote a generic constant the value of which may change at different places.

\vskip 0.5cm
{\bf Proof of Theorem 1.}  We only consider a one-period economy and omit the subscript index $t$. We consider a small perturbation of the tangency portfolio $r_{j,M}$ and start with a portfolio consisting of $r_j$, $r_{j,M}$, and $r_f$ with weights $\alpha$, $1$, and $-\alpha$, respective. The total wealth is $(\alpha+1-\alpha)=1$. Denote the new portfolio by $r_{\alpha}$ and it can be written as
\[r_{\alpha}=r_{j,M}+\alpha r_j-\alpha r_f.\]
The variance of $r_\alpha$ is 
\[\sigma_{\alpha}^2=\Var(r_{j,M}+\alpha r_j-\alpha r_f)=\sigma_{j,M}^2+2\alpha\gamma_{j,M}+\alpha^2\sigma_j^2,\]
where $\sigma_{j,M}^2=\Var(r_{j,M})$, $\sigma_j^2=\Var(r_j)$, and $\gamma_{j,M}=\Cov(r_{j,M},r_j)$. The expected return of $r_\alpha$ is 
\[\mu_\alpha=\mu_{j,M}+\alpha\mu_j-\alpha r_f.\]
It follows that
\[\frac{\partial \mu_\alpha}{\partial \alpha}=\mu_j-r_f,\]
and
\[\frac{\partial \sigma_\alpha}{\partial \alpha}=\frac{1}{2}(\sigma_{j,M}^2+2\alpha\gamma_{j,M}+\alpha^2\sigma_j^2)^{-1/2}(2\gamma_{j,M}+2\alpha\sigma_j^2).\]
At the tangency portfolio with $\alpha=0$, it is known from the mean-variance theory  that the slop of the capital allocation line (CAL) in Figure~\ref{fig00} is 
$\frac{\mu_{j,M}-r_f}{\sigma_{j,M}}$ as mentioned above.
On the other hand
\[\frac{\partial \mu_\alpha/\partial \alpha}{\partial \sigma_{\alpha}/\partial \alpha}|_{\alpha=0}=\frac{\mu_j-r_f}{\gamma_{j,M}/\sigma_{j,M}}.\]
From the mean-variance theory and the efficiency of the tangency portfolio $r_{j,M}$ on the frontier together with the complete market assumption as Definition ~\ref{def1} and Definition~\ref{def2}, we can conclude that the ratio between the partial derivatives above is equal to the slope of the capital allocation line:
\[\frac{\mu_{j,M}-r_f}{\sigma_{j,M}}=\frac{\mu_j-r_f}{\gamma_{j,M}/\sigma_{j,M}},\]
implying that
\[\mu_j-r_f=\frac{\gamma_{j,M}}{\sigma_{j,M}^2}(\mu_{j,M}-r_f)=\frac{\Cov(r_{j},\bw_{j}'\br)}{\Var(\bw_j'\br)}(\mu_{j,M}-r_f)=\rho_j(\mu_{j,M}-r_f),\]
where 
\[\rho_j=\frac{\Cov(r_{j},\bw_{j}'\br)}{\Var(\bw_j'\br)},\]
is the spatial rho associated with the $j$-th asset $r_j$. This completes the proof. $\Box$
\vskip 0.3cm
{\noindent\bf Proof of Theorem~\ref{thm1}.} By (\ref{ft:sp}), it follows that
\[\wh\bSigma_{yf}=\frac{1}{T}\sum_{t=1}^T\by_t\bff_t'=\bD(\boldsymbol{\rho})\bW\frac{1}{T}\sum_{t=1}^T\by_t\bff_t'+\bB\frac{1}{T}\sum_{t=1}^T\bff_t\bff_t'+\frac{1}{T}\sum_{t=1}^T\bve_t\bff_t'\]
and
\[\wh\bSigma_{yf}(1)=\frac{1}{T}\sum_{t=2}^T\by_t\bff_{t-1}'=\bD(\boldsymbol{\rho})\bW\frac{1}{T}\sum_{t=2}^T\by_t\bff_{t-k}'+\bB\frac{1}{T}\sum_{t=2}^T\bff_t\bff_{t-1}'+\frac{1}{T}\sum_{t=2}^T\bve_t\bff_{t-1}'.\]
Then,
\[\wh\bSigma_{yf}'\be_i=\wh\bSigma_{yf}'\bw_i\rho_i+\wh\bSigma_f\bb_i+\wh\bSigma_{\ve f}'\be_i,\]
and
\[\wh\bSigma_{yf}(1)'\be_i=\wh\bSigma_{yf}(1)'\bw_i\rho_i+\wh\bSigma_f(1)'\bb_i+\wh\bSigma_{\ve f}(1)'\be_i.\]
Therefore, we obtain that
\[\wh\bY_i=\left(\begin{array}{cc}
    \wh\bSigma_{yf}'\bw_i & \wh\bSigma_f' \\
    \wh\bSigma_{yf}(1)'\bw_i & \wh\bSigma_f(1)'
\end{array}\right)\bbeta_i+\left(\begin{array}{c}
    \wh\bSigma_{\ve f}'\be_i  \\
      \wh\bSigma_{\ve f}(1)'\be_i
\end{array}\right)=\wh\bX_i\bbeta_i+\left(\begin{array}{c}
    \wh\bSigma_{\ve f}'\be_i  \\
      \wh\bSigma_{\ve f}(1)'\be_i
\end{array}\right).\]
It follows that
\begin{equation}\label{behat:d}
\wh\bbeta_i(\lambda_i)=(\wh\bX_i(\lambda_i))^{-1}\wh\bX_i'\wh\bY_i=(\wh\bX_i(\lambda_i))^{-1}\wh\bX_i'\wh\bX_i\bbeta_i+(\wh\bX_i(\lambda_i))^{-1}\wh\bX_i'\left(\begin{array}{c}
    \wh\bSigma_{\ve f}'\be_i  \\
      \wh\bSigma_{\ve f}(1)'\be_i
\end{array}\right),
\end{equation}
where $\wh\bX_i(\lambda_i)=\wh\bX_i'\wh\bX_i+\lambda_i\bI_{K+1}$.
By Assumption~\ref{asm1} and a similar argument as that in (A.2) of the supplement of \cite{gao2022modeling}, we can show that 
\[\|\wh\bSigma_{yf}'(k)\bw_i-\bSigma_{yf}'(k)\bw_i\|_F=O_p(\sqrt{\frac{N}{T}}),\]
and
\[\|\wh\bSigma_f(k)-\bSigma_f(k)\|_F=O_p(\sqrt{\frac{1}{T}}),\]
where the first rate $\sqrt{N/T}$ can be reduced if some weak cross-sectional dependence is imposed.
Furthermore, by a similar argument, we can show that
\[\wh\bSigma_{\ve f}(k)'\be_i=\frac{1}{T}\sum_{t=k+1}^T\bff_{t-k}\ve_{i,t}=O_p(\sqrt{\frac{1}{T}}).\]
Therefore, if $N=o(T)$, by Assumption \ref{asm6}, we have
\[\|\wh\bbeta_i(\lambda_i)-\wh\bX_i(\lambda_i)^{-1}\wh\bX_i'\wh\bX_i\bbeta_i\|_2\leq C\|(\wh\bX_i(\lambda_i))^{-1}\wh\bX_i'\|_2\left\|\left(\begin{array}{c}
    \wh\bSigma_{\ve f}'\be_i  \\
      \wh\bSigma_{\ve f}(1)'\be_i
\end{array}\right)\right\|_2=O_p(T^{-1/2}),\]
and letting $\lambda_i\rightarrow 0$,
\[\|(\wh\bX_i'\wh\bX_i)(\wh\bbeta_i-\bbeta_i)\|_2\leq C\|\wh\bX_i'\|_2\left\|\left(\begin{array}{c}
    \wh\bSigma_{\ve f}'\be_i  \\
      \wh\bSigma_{\ve f}(1)'\be_i
\end{array}\right)\right\|_2=O_p(T^{-1/2}).\]
This completes the proof. $\Box$
\vskip 0.5 cm
{\noindent\bf Proof of Theorem~\ref{thm2}.} We only prove the case when $N$ is diverging in Theorem \ref{thm2}(ii) as the proof for (i) is similar. By (\ref{behat:d}),
\[(\wh\bX_i'\wh\bX_i)(\wh\bbeta_i-\bbeta_i)=\wh\bX_i'\left(\begin{array}{c}
   \frac{1}{T}\sum_{t=1}^T\bff_t\ve_{i,t} \\
      \frac{1}{T}\sum_{t=2}^T\bff_{t-1}\ve_{i,t}
\end{array}\right).   \]
To prove Theorem \ref{thm2}(ii), it is sufficient to show the following two statements,
\begin{equation}\label{vi}
    \wh\bX_i'\wh\bX_i\rightarrow_p \bV_i,
\end{equation}
and
\begin{equation}\label{dist:n}
   \sqrt{T}\left(\begin{array}{c}
   \frac{1}{T}\sum_{t=1}^T\bff_t\ve_{i,t} \\
     \frac{1}{T}\sum_{t=2}^T\bff_{t-1}\ve_{i,t}
\end{array}\right)\rightarrow_d N({\bf 0},\bU_i).
\end{equation}
By a similar argument as that in the proof of Theorem 1 above, we can show that
\[\wh\bX_i\rightarrow_p\bX_i:=\left(\begin{array}{cc}
    \bSigma_{yf}'\bw_i & \bSigma_f' \\
    \bSigma_{yf}(1)'\bw_i & \bSigma_f(1)'
\end{array}\right),\]
if $N=o(T)$. Therefore, we have
\[\wh\bX_i'\wh\bX_i\rightarrow_p\left(\begin{array}{cc}
     \bw_i'\bSigma_{yf}\bSigma_{yf}'\bw_i+\bw_i'\bSigma_{yf}(1)\bSigma_{yf}'(1)\bw_i &\bw_i'\bSigma_{yf}\bSigma_f\bw_i+\bw_i'\bSigma_{yf}(1)\bSigma_f'(1) \\
  \bSigma_f\bSigma_{yf}'\bw_i+\bSigma_f(1)\bSigma_{yf}'(1)\bw_i   &\bSigma_f^2+\bSigma_f(1)\bSigma_f'(1) 
\end{array}\right)=\bV_i.\]
By a similar argument as that in the proof of Theorem~\ref{thm1}, we can show that
\[\wh\bX_i'\wh\bX_i-\bX_i'\bX_i=O_p(N^{-1/2}T^{1/2}),\]
which implies (\ref{vi}) if $N=o(T)$.
To show (\ref{dist:n}), it is sufficient to prove that, for any vector $\ba=(\ba_1',\ba_2')'$ with $\ba_1\in R^K$ and $\ba_2\in R^K$, 
\begin{align}\label{linear:c}
    \sqrt{T}\ba'\left(\begin{array}{c}
   \frac{1}{T}\sum_{t=1}^T\bff_t\ve_{i,t} \\
      \frac{1}{T}\sum_{t=2}^T\bff_{t-1}\ve_{i,t}
\end{array}\right)\notag
=\ba_1'\frac{1}{\sqrt{T}}\sum_{t=1}^T\bff_t\ve_{i,t}+\ba_2'\frac{1}{\sqrt{T}}\sum_{t=1}^T\bff_{t-1}\ve_{i,t}\notag\\
\end{align}
is asymptotically normal.
Define
\begin{align}\label{snt}
    \bS_{N,T}=a_1\frac{1}{\sqrt{T}}\sum_{t=1}^T\bff_t\ve_{i,t}+\ba_2'\frac{1}{\sqrt{T}}\sum_{t=2}^T\bff_{t-1}\ve_{i,t},
\end{align}
we only need to show the asymptotic normality of $\bS_{N,T}$. By Schwarz's Inequality and Assumptions~\ref{asm2} and \ref{asm5}, we can derive that
\[E|\bff_t\ve_{i,t}|^{\gamma}\leq (E|\bff_t|^{2\gamma})^{1/2}(E|\ve_{it}|^{2\gamma})^{1/2}<\infty.\]
We now calculate the variance of $\bS_{N,T}$. Since it involves 16 terms in total and we start with the first term in (\ref{snt}). By definition and an elementary argument, we have
\[\Var(\frac{1}{\sqrt{T}}\sum_{t=1}^T\bff_t\ve_{i,t})=\bSigma_{f\ve_i}(0)+\sum_{j=1}^{T-1}(1-\frac{j}{T})\bSigma_{f\ve_i}(0,j).\]
Note that $\sum_{j=1}^\infty\alpha_N(j)^{1-2/\gamma}<\infty$ from Assumption~\ref{asm1}, by Proposition 2.5 of \cite{fan2003nonlinear}, we have
\begin{align*}
    \sup_i\sum_{j=1}^\infty|\bSigma_{f\ve_i}(0,j)|\leq C\sup_i\sum_{j=1}^\infty\alpha(j)^{1-2/\gamma}(E|\bff_t|^{2\gamma})^{1/\gamma}(E|\ve_{i,t}|^{2\gamma})^{1/\gamma}<\infty.
\end{align*}
We can calculate all the terms of $\bS_{N,T}$ and sum them up, by the Dominated Convergence  theorem, we have
\[\Var\left(\bS_{N,T}\right)\rightarrow \ba'\bU_i\ba.\]
To show the asymptotic normality of $\bS_{N,T}$, we employ the small-block and large-block techniques commonly used for weakly dependent data. Specifically, we partition the set $\{1,...,T\}$ into $2k_T+1$ subsets with large blocks of size $l_T$, small blocks of size $s_T$, and the last remaining set of size $T-k_T(l_T+s_T)$. Let
\[l_T=[\sqrt{T}/\log(T)],\,\,s_T=[\sqrt{T}\log(T)]^{\delta},\,\,k_T=[T/(l_T+s_T)],\]
where $[x]$ is the greatest integer less than or equal to $x$, and $1-2/\gamma\leq \delta<1$.
It is not hard to see that
\[l_T/\sqrt{T}\rightarrow 0,\,\,s_T/l_T\rightarrow 0,\,\,\text{and}\,\, k_T=O(\sqrt{T}\log(T)).\]
By Assumption~\ref{asm1} that $\sum_{j=1}^\infty\alpha_N(j)^{1-2/\gamma}$, we have $\alpha_N(s_T)=o(s_T^{-\gamma/(\gamma-2)})$. It follows that
\[k_T\alpha_N(s_T)=o(k_T/s_T^{\gamma/(\gamma-2)})=o(1).\]
Then,
we rewrite $\bS_{N,T}$ as
\begin{align}
   \bS_{N,T}=&\ba_1'\frac{1}{\sqrt{T}}\sum_{j=1}^{k_T}\bxi_{j}^{(1)}+\ba_2' \frac{1}{\sqrt{T}}\sum_{j=1}^{k_T}\bxi_{j}^{(2)}
   +\ba_1'\frac{1}{\sqrt{T}}\sum_{j=1}^{k_T}\bfeta_{j}^{(1)}+\ba_2' \frac{1}{\sqrt{T}}\sum_{j=1}^{k_T}\bfeta_{j}^{(2)}\notag\\
   &+\ba_1'\frac{1}{\sqrt{T}}\bzeta_{j}^{(1)}+\ba_2' \frac{1}{\sqrt{T}}\bzeta_{j}^{(2)},
\end{align}
where
\[\bxi_j^{(1)}=\sum_{t=(j-1)(l_T+s_T)+1}^{jl_T+(j-1)s_T}\bff_t\ve_{i,t},\,\,\bfeta_j^{(1)}=\sum_{t=jl_T+(j-1)s_T+1}^{j(l_T+s_T)}\bff_t\ve_{i,t},\]
\[\bzeta_j^{(1)}=\sum_{t=k_T(l_T+s_T)+1}^{T}\bff_t\ve_{i,t},\,\,\bxi_j^{(2)}=\sum_{t=(j-1)(l_T+s_T)+1}^{jl_T+(j-1)s_T}\bff_{t-1}\ve_{i,t},\]
\[\bfeta_j^{(2)}=\sum_{t=jl_T+(j-1)s_T+1}^{j(l_T+s_T)}\bff_{t-1}\ve_{i,t},\,\,\bzeta_j^{(2)}=\sum_{t=k_T(l_T+s_T)+1}^{T}\bff_{t-1}\ve_{i,t},\]

Note that $\alpha_N(T)=o(T^{2/\gamma-1})$, $k_Ts_T/T\rightarrow 0$, and $(l_T+s-T)/T\rightarrow 0$, it follows from Proposition 2.7 of \cite{fan2003nonlinear} that
\[\frac{1}{\sqrt{T}}\sum_{j=1}^{k_T}\bfeta_j^{(l)}=o_p(1),\,\,\text{and}\,\,\frac{1}{\sqrt{T}}\bzeta_j^{(l)}=o_p(1),\,\,l=1,2,3,4.\]
Then,
\[\bS_{N,T}=\ba_1'\frac{1}{\sqrt{T}}\sum_{j=1}^{k_T}\bxi_{j}^{(1)}+\ba_2' \frac{1}{\sqrt{T}}\sum_{j=1}^{k_T}\bxi_{j}^{(2)}+o_p(1).\]
By a similar argument as Theorem 2.21 of \cite{fan2003nonlinear}, we can show that 
\[\sqrt{T}\ba'\left(\begin{array}{c}
   \frac{1}{T}\sum_{t=1}^T\bff_t\ve_{i,t} \\
      \frac{1}{T}\sum_{t=2}^T\bff_{t-1}\ve_{i,t}
\end{array}\right)\longrightarrow_d N({ 0},\ba'\bU_i\ba).\]
We replace $\ba$ by $(\bU_i^{-1/2})'\ba$ and obtain
\[\sqrt{T}\ba'\bU_i^{-1/2}\left(\begin{array}{c}
   \frac{1}{T}\sum_{t=1}^T\bff_t\ve_{i,t} \\
      \frac{1}{T}\sum_{t=2}^T\bff_{t-1}\ve_{i,t}
\end{array}\right)\longrightarrow_d N({ 0},1),\]
which implies that
\begin{equation}\label{asym:n}
\sqrt{T}\bU_i^{-1/2}\left(\begin{array}{c}
  \frac{1}{T}\sum_{t=1}^T\bff_t\ve_{i,t} \\
      \frac{1}{T}\sum_{t=2}^T\bff_{t-1}\ve_{i,t}
\end{array}\right)\longrightarrow_d N({\bf 0},\bI_{2K}).
\end{equation}
Therefore, 
\begin{equation}\label{asym:full}
    \sqrt{T}\bV_i(\wh\bbeta_i-\bbeta_i)\longrightarrow_d N({\bf 0}, \bX_i'\bU_i\bX_i).
\end{equation}
Theorem \ref{thm2} follows from (\ref{vi}) and (\ref{asym:full}). This completes the proof. $\Box$

\vskip 0.5cm

{\noindent\bf Proof of Theorem \ref{thm3}.} Note that
\begin{align}\label{sig:y}
    \wh\bSigma_y(k)=&\frac{1}{T}\sum_{t=k+1}^T\by_t\by_{t-k}'\notag\\
    =&\frac{1}{T}\sum_{t=k+1}^T\{\bLambda\bff_t\bff_{t-k}'\bLambda'+\bLambda\bff_t\bxi_{t-k}'+\bxi_t\bff_{tk}'\bLambda'+\bxi_t\bxi_{t-k}'\}\notag\\
    =&\bLambda\frac{1}{T}\sum_{t=k+1}^T(\bff_t\bff_{t-k}'\bLambda'+\bff_t\bxi_{t-k}')+\frac{1}{T}\sum_{t=k+1}^T(\bxi_t\bff_{t-k}'\bLambda'+\bxi_t\bxi_{t-k}')\notag\\
    =&\bLambda\bG_{1,k}+\bG_{2,k},
\end{align}
where
\[\bG_{1,k}=\frac{1}{T}\sum_{t=k+1}^T(\bff_t\bff_{t-k}'\bLambda'+\bff_t\bxi_{t-k}'),\,\,\bG_{2,k}=\frac{1}{T}\sum_{t=k+1}^T(\bxi_t\bff_{t-k}'\bLambda'+\bxi_t\bxi_{t-k}').\]
It follows from the definition of $\wh\bM$ in (\ref{mhat}) that 
\begin{align}\label{Mhat:exp}
    \wh\bM=&\sum_{k=1}^{k_0}\wh\bSigma_y(k)\wh\bSigma_y'(k)\notag\\
    =&\sum_{k=1}^{k_0}(\bLambda\bG_{1,k}+\bG_{2,k})(\bLambda\bG_{1,k}+\bG_{2,k})'\notag\\
    =&\bLambda\sum_{k=1}^{k_0}\bG_{1,k}\bG_{1,k}'\bLambda'+\sum_{k=1}^{k_0}(\bLambda\bG_{1,k}\bG_{2,k}'+\bG_{2,k}\bG_{1,k}'\bLambda'+\bG_{2,k}\bG_{2,k}').
\end{align}
Let $\wh\bV_{NT}\in R^{r}$ be a diagonal matrix with diagonal elements being the top $K$ eigenvalues of $\wh\bM$, it follows from Assumptions~\ref{asm3} and \ref{asm4} that $\wh\bV_{NT}\asymp O(N^2)$. Since the columns of $\wh\bLambda$ are the eigenvectors of $\wh\bM$, it follows that
\[\wh\bM\wh\bLambda=\wh\bLambda\wh\bV_{NT},\]
implying that
\begin{align}\label{lbd:hat}
    \wh\bLambda=&\wh\bM\wh\bLambda\wh\bV_{NT}^{-1}\notag\\
=&\bLambda\sum_{k=1}^{k_0}\bG_{1,k}\bG_{1,k}'\bLambda'\wh\bLambda\wh\bV_{NT}^{-1}+\sum_{k=1}^{k_0}\left[\bLambda\bG_{1,k}\bG_{2,k}'+\bG_{2,k}\bG_{1,k}'\bLambda'+\bG_{2,k}\bG_{2,k}'\right]\wh\bLambda\wh\bV_{NT}^{-1}.
\end{align}
Let $\bH_{NT}'=\sum_{k=1}^{k_0}\bG_{1,k}\bG_{1,k}'\bLambda'\wh\bLambda\wh\bV_{NT}^{-1}$, it follows that $\bH= O_p(1)$ and $\bH^{-1}=O_p(1)$. Then (\ref{lbd:hat}) implies that
\begin{equation}\label{labd:diff}
    \wh\bLambda-\bLambda\bH_{NT}'=\sum_{k=1}^{k_0}\left[\bLambda\bG_{1,k}\bG_{2,k}'+\bG_{2,k}\bG_{1,k}'\bLambda'+\bG_{2,k}\bG_{2,k}'\right]\wh\bLambda\wh\bV_{NT}^{-1}.
\end{equation}
First, by Assumption~\ref{asm1} and a similar argument as that in (A.2) of the supplement of \cite{gao2022modeling}, we can show that
\[\|\bG_{1,k}\|_F=\|\frac{1}{T}\sum_{t=k+1}^T(\bff_t\bff_{t-k}'\bLambda'+\bff_t\bxi_{t-k}')\|_F=O_p(\sqrt{N})+O_p(1+\sqrt{\frac{N}{T}})=O_p(\sqrt{N}),\]
and
\[\|\bG_{2,k}\|_F=\|\frac{1}{T}\sum_{t=k+1}^T(\bxi_t\bff_{t-k}'\bLambda'+\bxi_t\bxi_{t-k}')\|_F=O_p(\sqrt{\frac{N}{T}}\sqrt{N})+O_p(\sqrt{\frac{N^2}{T}})=O_p(\sqrt{\frac{N^2}{T}}).\]
Then, it follows from (\ref{labd:diff}) and the above rates that
\[\|\wh\bLambda-\bLambda\bH'\|_F=O_p(\sqrt{N}\sqrt{N}\sqrt{\frac{N^2}{T}}+\sqrt{N}\sqrt{N}\sqrt{\frac{N^2}{T}}+\frac{N^2}{T})O_p(\sqrt{N}/N^2)=O_p(\sqrt{\frac{N}{T}}),\]
implying that
\[\frac{1}{\sqrt{N}}\|\wh\bLambda-\bLambda\bH'\|_F=O_p(T^{-1/2}).\]
This completes the proof of Theorem \ref{thm3}. $\Box$

\vskip 0.5cm

\begin{lemma}\label{lm1}
    Let Assumptions~\ref{asm1}--\ref{asm8} hold. Then, as $N,T\rightarrow\infty$,
    \[\bH_{NT}\bH_{NT}'=\bI_K+O_p(T^{-1/2}),\,\,\text{and}\,\,
    \bH_{NT}'\bH_{NT}=\bI_K+O_p(T^{-1/2}).\]
\end{lemma}
{\bf Proof.} First, note that
\begin{align}
 \bH_{NT}\bH_{NT}'-\bI_K=&\bH_{NT}\frac{\bLambda'\bLambda}{N}\bH_{NT}'-\frac{\wh\bLambda'\wh\bLambda}{N}\notag\\
 =&\frac{1}{\sqrt{N}}(\bH_{NT}\bLambda'-\wh\bLambda')\frac{1}{\sqrt{N}}\bLambda\bH_{NT}'+\frac{1}{\sqrt{N}}\wh\bLambda'(\bLambda\bH_{NT}'-\wh\bLambda)/\sqrt{N}.
\end{align}
Then, it follows from Theorem 3 that
\[\|\bH_{NT}\bH_{NT}'-\bI_K\|_F=O_p(T^{-1/2}).\]
Furthermore, since $\bH_{NT}=O_p(1)$ and $\bH_{NT}^{-1}=O_p(1)$, then
\[\bH_{NT}'\bH_{NT}\bH_{NT}'=\bH_{NT}'+O_p(T^{-1/2}),\]
we multiply $\bH_{NT}^{-1}$ on the right of both sides and obtain
\[\bH_{NT}'\bH_{NT}=\bI_K+O_p(T^{-1/2}).\]
This completes the proof. $\Box$

\begin{lemma}
    Let Assumptions~\ref{asm1}--\ref{asm8} hold. Then, as $N,T\rightarrow\infty$,
    \[\wh\bLambda'\wh\bM\wh\bLambda=\wh\bV_{NT}\rightarrow_p\bV,\]
    where $\bV$ is a diagonal matrix consisting of the top $K$ eigenvalues of $\bM$ defined in (\ref{M}).
\end{lemma}
{\bf Proof.} The proof is similar to Theorem 1 of \cite{lam2012factor}. We omit the details to save space. $\Box$
\begin{lemma}
    Let Assumptions~\ref{asm1}--\ref{asm8} hold. Then there exists an orthogonal matrix $\bH\in R^K$ such that
    $\bH_{NT}\rightarrow_p\bH$ with probability tending to one as $N,T\rightarrow\infty$.
\end{lemma}
{\bf Proof.} Note that
\[\bG_{1,k}=\wh\bSigma_f(k)\bLambda'+\wh\bSigma_{f\xi}(k).\]
If $N=o(T)$, we have that $\wh\bSigma_f(k)\rightarrow_p\bSigma_f(k)$ and $\wh\bSigma_{f\xi}(k)\rightarrow_p\bSigma_{f\xi}(k)$.
By definition,
\[\bH_{NT}'=\sum_{k=1}^{k_0}\bG_{1,k}\bG_{1,k}'\bLambda'\wh\bLambda\wh\bV_{NT}^{-1}=\sum_{k=1}^{k_0}\bG_{1,k}\bG_{1,k}'\bLambda'\bLambda\bH_{NT}'\bV^{-1}+o_p(1).\]
Then, by Lemma~\ref{lm1},
\begin{align}
 \bH_{NT}'\frac{\bV}{N^2}\bH_{NT}=&\frac{1}{N}\sum_{k=1}^{k_0}(\bSigma_f(k)\bLambda'+\bSigma_{f\xi}(k)))(\bSigma_f(k)\bLambda'+\bSigma_{f\xi}(k)))'+o_p(1)\notag\\
 =&\sum_{k=1}^{k_0}\bSigma_f(k)\bSigma_f'(k)+o_p(1).
\end{align}
Therefore, $\bH_{NT}$ will converge to the matrix consisting of the eigenvectors of $\sum_{k=1}^{k_0}\bSigma_f(k)\bSigma_f'(k)$, denoted by $\bH$. This completes the proof. $\Box$
\vskip 0.5cm
{\noindent\bf Proof of Theorem \ref{thm4}.} We first show Theorem \ref{thm4}(i). Note that

\begin{align}
    \wh\bff_t=&\frac{1}{N}\wh\bLambda'\by_t\notag\\
    =&\frac{1}{N}\wh\bLambda'(\bLambda\bff_t+\bxi_t)\notag\\
    =&\frac{1}{N}\wh\bLambda'\bLambda\bff_t+\frac{1}{N}\wh\bLambda'\bxi_t\notag\\
    =&\bK_{NT}\bff_t+\frac{1}{N}(\wh\bLambda-\bLambda\bH_{NT}')'\bxi_t+\frac{1}{N}\bH_{NT}\bLambda'\bxi_t,
\end{align}
where $\bK_{NT}=\frac{1}{N}\wh\bLambda'\bLambda$ which has the same limit $\bH$ as that of $\bH_{NT}$, but they are not identical in finite samples. 
By Assumption~\ref{asm7}, it is not hard to show that
\[\max_{1\leq t\leq t}|f_{i,t}|=O_p(\log(T)),\,\,\text{and}\,\,\max_{1\leq t\leq T}|\ve_{i,t}|=O_p(\log(T)).\]
Therefore,
\[\max_{1\leq t\leq T}\|\frac{1}{N}(\wh\bLambda-\bLambda\bH_{NT}')'\bxi_t\|_F\leq \frac{1}{\sqrt{N}}\|\wh\bLambda-\bLambda\bH_{NT}'\|_F\max_{1\leq t\leq T}\|\frac{\bxi_t}{\sqrt{N}}\|_F=O_p(T^{-1/2}\log(T)),\]
and
\[\max_{1\leq t \leq T}\|\frac{1}{N}\bH_{NT}\bLambda'\bxi_t\|_F\leq C\frac{1}{\sqrt{N}}\|\bH_{NT}\|_F\max_{1\leq t\leq T}\|\frac{1}{\sqrt{N}}\bLambda'\bxi_t\|_F=O_p(N^{-1/2}\log(T)).\]
Then, it follows from the above rates that
\[\max_{1\leq t\leq t}\|\wh\bff_t-\bK_{NT}\bff_t\|_F=O_p((\frac{1}{\sqrt{T}}+\frac{1}{\sqrt{N}})\log(T)).\]
This proves Theorem \ref{thm4}(i). 

For Theorem \ref{thm4}(ii), if $N=o(T)$,
\begin{equation}\label{fhat}
  \sqrt{N}(\wh\bff_t-\bK_{NT}\bff_t)=\bH\frac{1}{\sqrt{N}}\bLambda'\bxi_t+o_p(1).
\end{equation}
By Assumption~\ref{asm8}, we have
\[\sqrt{N}(\wh\bff_t-\bK_{NT}\bff_t)\longrightarrow_d N(0,\bH\bGamma_t\bH').\]
This completes the proof. $\Box$
\vskip 0.5cm

{\bf Proof of Theorem \ref{thm5}.} We use $\bH$ instead of $\bH_{NT}$ for simplicity in this proof. By definition,
\begin{align}
    \wt\bSigma_{yf}(k)=&\frac{1}{T}\sum_{t=k+1}^T\by_t\wh\bff_{t-k}'\notag\\
    =&\frac{1}{T}\sum_{t=k+1}^T\bD(\boldsymbol{\rho})\bW\by_t\wh\bff_{t-k}'+\bB\frac{1}{T}\sum_{t=k+1}^T\bff_t\wh\bff_{t-k}'+\frac{1}{T}\sum_{t=k+1}^T\bve_t\wh\bff_{t-k}'\notag\\
    =&\bD(\boldsymbol{\rho})\bW\wt\bSigma_{yf}(k)+\bB\bK_{NT}^{-1}\wt\bSigma_{f}(k)+\bB\frac{1}{T}\sum_{t=k+1}^T(\bff_t-\bK_{NT}^{-1}\wh\bff_t)\wh\bff_{t-k}'+\frac{1}{T}\sum_{t=k+1}^T\bve_t\wh\bff_{t-k}'.
\end{align}
Then, it follows that
\begin{equation}\label{sigk:t}
\wt\bSigma_{yf}(k)'\be_i=\wt\bSigma_{yf}(k)'\bw_i\rho_i+\wt\bSigma_f(k)'(\bK_{NT}')^{-1}\bb_i+\frac{1}{T}\sum_{t=k+1}^T\wh\bff_{t-k}(\bff_t-\bK_{NT}^{-1}\wh\bff_t)'\bb_i+\frac{1}{T}\sum_{t=k+1}^T\wh\bff_{t-k}\ve_{i,t}.  
\end{equation}
We now analyze the last two terms. First,
\begin{align}
 \frac{1}{T}\sum_{t=k+1}^T\wh\bff_{t-k}(\bff_t-\bK_{NT}^{-1}\wh\bff_t)'\bb_i=&\frac{1}{T}\sum_{t=k+1}^T(\wh\bff_{t-k}-\bK_{NT}\bff_{t-k})(\bK_{NT}\bff_t-\wh\bff_t)'(\bK_{NT}')^{-1}\bb_i\notag\\
 &+\frac{1}{T}\sum_{t=k+1}^T\bK_{NT}\bff_{t-k}(\bK_{NT}\bff_t-\wh\bff_t)'(\bK_{NT}')^{-1}\bb_i,
\end{align}
where
\[\|\frac{1}{T}\sum_{t=k+1}^T(\wh\bff_{t-k}-\bK_{NT}\bff_{t-k})(\bK_{NT}\bff_t-\wh\bff_t)'(\bK_{NT}')^{-1}\bb_i\|_2=O_p((\frac{1}{T}+\frac{1}{N})\log(T)^2).\]
By (\ref{fhat}), and $\bff_{t-k}$ and $\bxi_t$ are uncorrelated,  we  have
\begin{align}\label{KFK}
 \frac{1}{T}\sum_{t=k+1}^T\bK_{NT}\bff_{t-k}(\bK_{NT}\bff_t-\wh\bff_t)'(\bK_{NT}')^{-1}\bb_i=&\frac{1}{T}\sum_{t=k+1}^T\bK_{NT}\bff_{t-k}\bxi_t'(\bLambda\bH_{NT}'-\wh\bLambda)(\bK_{NT})^{-1}\bb_i/N\notag\\
 &- \frac{1}{T}\sum_{t=k+1}^T\bK_{NT}\bff_{t-k}\bxi_t'\bLambda\bH_{NT}'(\bK_{NT})^{-1}\bb_i/N\notag\\
 =&O_p(\frac{1}{T}+\frac{1}{\sqrt{NT}}).
\end{align}
Next, we consider the last term of (\ref{sigk:t}).
\[\frac{1}{T}\sum_{t=k+1}^T\wh\bff_{t-k}\ve_{i,t}=\frac{1}{T}\sum_{t=k+1}^T\bK_{NT}\bff_{t-k}\ve_{i,t}+\frac{1}{T}\sum_{t=k+1}^T(\wh\bff_{t-k}-\bK_{NT}\bff_{t-k})\ve_{i,t}.\]
By a similar argument as (\ref{KFK}), we can show that
\[\|\frac{1}{T}\sum_{t=k+1}^T(\wh\bff_{t-k}-\bK_{NT}\bff_{t-k})\ve_{i,t}\|_F=O_p(\frac{1}{T}+\frac{1}{\sqrt{NT}}).\]
Then it follows from (\ref{sigk:t}) that
\[\wt\bY_i=\wt\bX_i\bK_{NT}^*\bbeta_i+\left(\begin{array}{c}
   \bK_{NT}\frac{1}{T}\sum_{t=1}^T\bff_t\ve_{i,t}  \\
      \bK_{NT}\frac{1}{T}\sum_{t=2}^T\bff_{t-1}\ve_{i,t} 
\end{array}\right)+O_p(\frac{1}{\sqrt{NT}}+(\frac{1}{T}+\frac{1}{N})\log(T)^2),\]
where $\bK_{NT}^*=\diag(1,\bK_{NT})$ and $\bbeta_i=(\rho_i,\bb_i')'$. Then,
\begin{equation}\label{beta:tl}
\wt\bbeta_i(\lambda_i)=\wt\bX_i(\lambda_i)^{-1}\wt\bX_i'\wt\bX_i\bK_{NT}^{*}\bbeta_i+\wt\bX_i(\lambda_i)^{-1}\wt\bX_i'\left(\begin{array}{c}
   \bK_{NT}\frac{1}{T}\sum_{t=1}^T\bff_t\ve_{i,t}  \\
      \bK_{NT}\frac{1}{T}\sum_{t=2}^T\bff_{t-1}\ve_{i,t} 
\end{array}\right)+\bR_i,  
\end{equation}
where $\bR_i$ is the remaining term, and we will show that $\sqrt{T}\bR_i=o_p(1)$. Note that
\begin{align}
  \wt\bSigma_{yf}(k)=\frac{1}{T}\sum_{t=k+1}^T\by_{t}\wh\bff_{t-k}'=&\frac{1}{T}\sum_{t=k+1}^T\by_{t}\bff_{t-k}'\bK_{NT}'+\frac{1}{T}\sum_{t=k+1}^T\by_{t}(\wt\bff_{t-k}-\bK_{NT}\bff_t)'\notag\\
  =&\wh\bSigma_{yf}\bK_{NT}'+O_p(N^{-1/2}+T^{-1/2})\notag\\
  \rightarrow_p&\wh\bSigma_{yf}\bK_{NT}',
\end{align}
if $N=o(T)$. Similarly, we can show that
\[\wt\bSigma_{f}(k)=\bH\bSigma_f(k)\bH'+o_p(1).\]
Therefore, if $N=o(T)$, 
\[\wt\bX_i=\left(\begin{array}{cc}
    \wt\bSigma_{yf}'\bw_i & \wt\bSigma_f' \\
    \wt\bSigma_{yf}'(1)\bw_i & \wt\bSigma_f'(1)
\end{array}\right)\rightarrow_p\bX_i^H=\left(\begin{array}{cc}
    \bH\bSigma_{yf}'\bw_i & \bH\bSigma_f\bH' \\
    \bH\bSigma_{yf}'(1)\bw_i & \bH\bSigma_f'(1)\bH'
\end{array}\right),\]
and hence
\[\wh\bX_i'\wt\bX_i\rightarrow_p\left(\begin{array}{cc}
     \bw_i'\bSigma_{yf}\bSigma_{yf}'\bw_i+\bw_i'\bSigma_{yf}(1)\bSigma_{yf}'(1)\bw_i &\bw_i'\bSigma_{yf}\bSigma_f\bH'+\bw_i'\bSigma_{yf}(1)\bSigma_f'(1)\bH' \\
  \bH\bSigma_f\bSigma_{yf}'\bw_i+\bH\bSigma_f(1)\bSigma_{yf}'(1)\bw_i   &\bH\bSigma_f^2\bH'+\bH\bSigma_f(1)\bSigma_f'(1)\bH' 
\end{array}\right)=\bV_i^H.\]
It follows that
\[(\wt\bX_i'\wt\bX_i+\lambda_i\bI_{K+1})^{-1}\wt\bX_i=O_p(1),\]
implying that $\sqrt{T}\bR_i=o_p(1)$. Then (\ref{beta:tl}) implies that
\[\wt\bbeta_i(\lambda_i)-\wt\bX_i(\lambda_i)^{-1}\wt\bX_i'\wt\bX_i\bK_{NT}^{*}\bbeta_i=O_p(T^{-1/2}).\]
Let $\lambda_i\rightarrow 0$, we obtain that
\[\sqrt{T}(\wt\bX_i'\wt\bX_i)(\wt\bbeta_i-\bK_{NT}^*\bbeta_i)=\wt\bX_i'\left(\begin{array}{c}
   \bK_{NT}\frac{1}{\sqrt{T}}\sum_{t=1}^T\bff_t\ve_{i,t}  \\
      \bK_{NT}\frac{1}{\sqrt{T}}\sum_{t=2}^T\bff_{t-1}\ve_{i,t}
\end{array}\right) +o_p(1).\]
By a similar argument as that in the proof of Theorem \ref{thm2}, we have
\begin{equation}\label{asym:ltf}
\sqrt{T}\left(\begin{array}{c}
   \bK_{NT}\frac{1}{T}\sum_{t=1}^T\bff_t\ve_{i,t} \\
      \bK_{NT}\frac{1}{T}\sum_{t=2}^T\bff_{t-1}\ve_{i,t}
\end{array}\right)\longrightarrow_d N({\bf 0},{\bU_i^H}),
\end{equation}
where 
\[\Var\left(\begin{array}{c}
   \bK_{NT}\frac{1}{\sqrt{T}}\sum_{t=1}^T\bff_t\ve_{i,t}  \\
      \bK_{NT}\frac{1}{\sqrt{T}}\sum_{t=2}^T\bff_{t-1}\ve_{i,t}
\end{array}\right)\rightarrow \bU_i^H,\]
which is defined as
\[\bU_i^H=\left(\begin{array}{cc}
     \bH\bSigma_{f\ve_i}(0)\bH' &\bH\bSigma_{f\ve_i}(1)\bH'\\
  \bH\bSigma_{f\ve_i}'(1)\bH' &\bH\bOmega_{f\ve_i}(0)\bH'
\end{array}\right),\]
where
 $\bSigma_{f\ve_i}(0)$, $\bSigma_{f\ve_i}(1)$, and $\bOmega_{f\ve_i}(0)$ are defined in Section \ref{sec3}. It follows from (\ref{asym:ltf}) that
    \[\sqrt{T}\bV_i^H(\wh\bbeta_i-\bK_{NT}^*\bbeta_i)\longrightarrow_d N({\bf 0}, \bX_i^H{'}\bU_i^{H}\bX_i^H).\]
 This completes the proof. $\Box$

\end{document}